\documentclass[12pt,twoside,a4paper,normalheadings,headsepline]{scrbook}

\usepackage{tdr_master_exe}
\usepackage{tdrdefinitions}
\providecommand{\href}[2]{\hspace*{-3mm}{}}

\begin{document}

\setcounter{part}{1}
\setcounter{chapter}{0}
\setcounter{secnumdepth}{3}
\setcounter{tocdepth}{2}
\addtolength{\parskip}{3mm}

\thispagestyle{empty}
\pagestyle{empty}
\begin{picture}(150,180)\unitlength 1mm

{\sfb
\put(0,40){\Huge TESLA}
\put(0,15){\begin{minipage}{14cm}
       \LARGE\bf
        \begin{flushleft}{\sfb\LARGE
        The Superconducting
        Electron Positron \\Linear Collider
        with an Integrated\\
        X-Ray Laser Laboratory\\
        }\end{flushleft}\end{minipage}}
\put(0,-10){\Huge Technical Design Report}
\put(60,-40){\LARGE Part I Executive Summary}

\put(-5,-150){{\sfb\bf\large DESY-2001-011, ECFA-2001-209}}
\put(-5,-155){{\sfb\bf\large TESLA-2001-23, TESLA-FEL-2001-05}}

\put(131,-150){\sfb \large March}
\put(135,-155){\sfb \large 2001}}
\end{picture}

\newpage

\thispagestyle{empty}

\begin{picture}(150,180)\unitlength 1mm
  \put(-10,-65){\includegraphics[height=2.5cm]{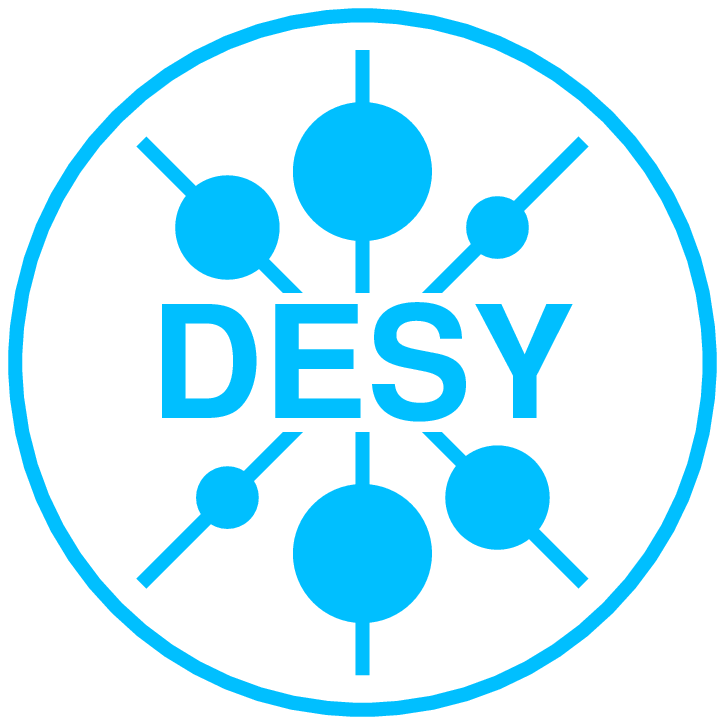}}
  \put(-10,-120){\begin{minipage}{10cm}{
   {\bf\noindent Publisher:} \\
   \noindent DESY\\
   \noindent Deutsches Elektronen-Synchroton\\
   \noindent Notkestra{\ss}e 85, D-22607 Hamburg\\
   \noindent Germany\\
   \noindent \url{http://www.desy.de}\\
   \noindent E-mail: desyinfo@desy.de\\

   \noindent Member of the Hermann von Helmholtz Association\\
   \noindent of National Research Centers (HGF)\\

   \noindent Reproduction including extracts is permitted \\
   \noindent subject to crediting the source.\\


   {\bf \noindent Copy deadline:} March 2001 \\
   ISBN 3-935702-00-0\\
   ISSN 0418-9833
}
\end{minipage}}

\end{picture}

\clearpage

\begin{center}
\begin{picture}(150,180)\unitlength 1mm
  \put(-50,-100){\includegraphics[height=10cm]{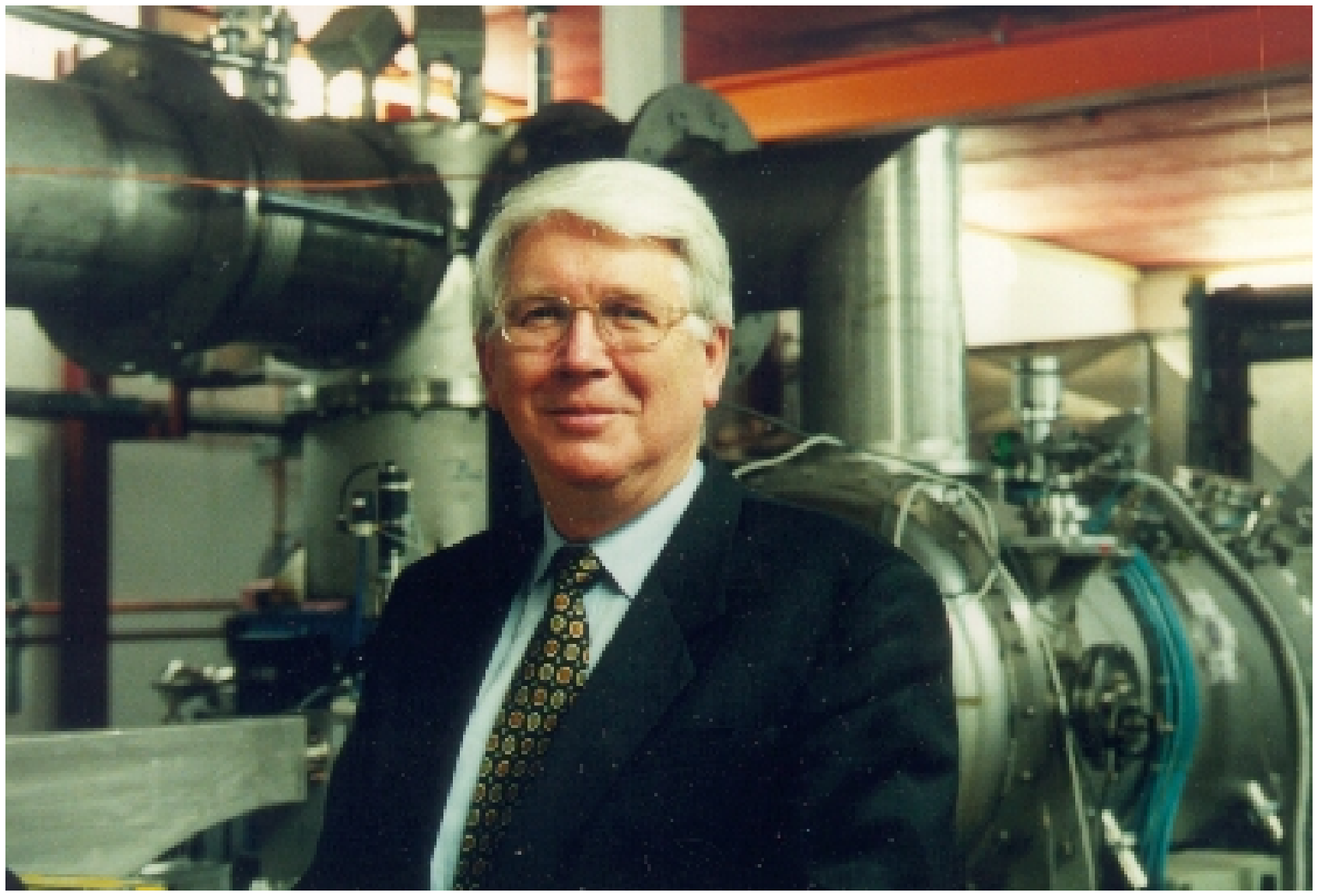}}
  \put(-20,-140){Dedicated to the memory of Bj{\o}rn H. Wiik (1937-1999)}
\end{picture}
\end{center}
\cleardoublepage

\pagestyle{empty}
\begin{picture}(150,180)\unitlength 1mm

{\sfb
\put(0,40){\Huge TESLA}
\put(0,15){\begin{minipage}{14cm}
       \LARGE\bf
        \begin{flushleft}{\sfb\LARGE
        The Superconducting
        Electron Positron \\Linear Collider
        with an Integrated\\
        X-Ray Laser Laboratory\\
        }\end{flushleft}\end{minipage}}
\put(0,-10){\Huge Technical Design Report}

\put(-5,-150){{\sfb\bf\large DESY-2001-011, ECFA-2001-209}}
\put(-5,-155){{\sfb\bf\large TESLA-2001-23, TESLA-FEL-2001-05}}

\put(131,-150){\sfb \large March}
\put(135,-155){\sfb \large 2001}}
\end{picture}

\newpage
\thispagestyle{empty}
{\large
\begin{picture}(150,180)\unitlength 1mm
\put(-5,20)   {{\bf\Large PART I: Executive Summary }}
      \put(25,12){Editors: \begin{minipage}[t]{10cm}{F.Richard, J.R.Schneider, D.Trines, A.Wagner}\end{minipage}}
\put(-5,-10) {{\bf \Large PART II: The Accelerator}}
      \put(25,-18){Editors: \begin{minipage}[t]{9cm}{R.Brinkmann,
      K.Fl\"ottmann, J.Rossbach,\hfil P.Schm\"user, N.Walker, H.Weise}\end{minipage}}
\put(-5,-40) {{\bf \Large PART III: Physics at an e$^+$e$^-$ Linear Collider}}
      \put(25,-48){Editors: \begin{minipage}[t]{10cm}{R.D.Heuer,
      D.Miller, F.Richard, P.Zerwas}\end{minipage} }
\put(-5,-70) {{\bf \Large PART IV: A Detector for TESLA}}
      \put(25,-78){Editors: \begin{minipage}[t]{10cm}{T.Behnke, S.Bertolucci,
      R.D.Heuer, R.Settles }\end{minipage}}
\put(-5,-100) {{\bf \Large PART V: The X-Ray Free Electron Laser Laboratory}}
      \put(25,-108){Editors:
      \begin{minipage}[t]{10cm}{G.Materlik, T.Tschentscher}\end{minipage}}
\put(-5,-130) {{\bf \Large PART VI: Appendices}}
      \put(25,-138){Editors: \begin{minipage}[t]{10cm}{
        R.Klanner\\
        Chapter 1: V.Telnov\\
        Chapter 2: U.Katz, M.Klein, A.Levy\\
        Chapter 3: R.Kaiser, W.D.Nowak\\
        Chapter 4: E.DeSanctis, J.-M.Laget, K.Rith
}\end{minipage}}
\end{picture}
}
\clearpage

\begin{picture}(150,180)\unitlength 1mm

  \put(20,50){\sfb \LARGE PART I: Executive Summary}
  \put(20,42){\sfb \large Editors:}
  \put(20,34){\sfb \large F. Richard, J.R.Schneider,}
  \put(20,26){\sfb \large D.Trines, A.Wagner}

\end{picture}

\clearpage

\cleardoublepage
\pagestyle{headings}
\providecommand{\totalcost}{3136}
\providecommand{\operating}{120}
\providecommand{\linaccost}{}
\providecommand{\xfelacccost}{241}
\providecommand{\xfellabcost}{290}
\providecommand{\detcostone}{160}
\providecommand{\detcosttwo}{280}

\providecommand{\xfelcost}{531}

\providecommand{\personyears}{6933}
\providecommand{\manpower}{xxx}

\providecommand{\gradientpercentage}{8}

\providecommand{\linea}{1131}
\providecommand{\lineb}{587}
\providecommand{\linec}{97}
\providecommand{\lined}{215}
\providecommand{\linee}{101}
\providecommand{\linef}{546}
\providecommand{\lineg}{336}
\providecommand{\lineh}{124}

\providecommand{\nfigure}[1] {#1}

\pagenumbering{roman}
\setcounter{page}{1}
\renewcommand\thefigure{\arabic{figure}}
\addchap*{TESLA -- A Summary}

This report describes the scientific aims and potential as well 
as the technical design of TESLA 
(\TeV--Energy Superconducting Linear Accelerator), 
a superconducting electron--positron collider of initially 500\,\GeV\  
total energy, extendable to 800\,\GeV, and an integrated X--ray laser
laboratory. A large--scale interdisciplinary and 
international research
campus will be created around TESLA to provide unique research 
possibilities for particle physics, for condensed matter physics, 
chemistry and material science, and for structural biology.
In this way TESLA satisfies the criteria for
new large endeavours in science: they should be unique, open completely new
research possibilities and should carry the promise to advance our
knowledge of nature in many branches of science. 
\begin{figure}[h!]
\begin{center}
\includegraphics[height=10cm]{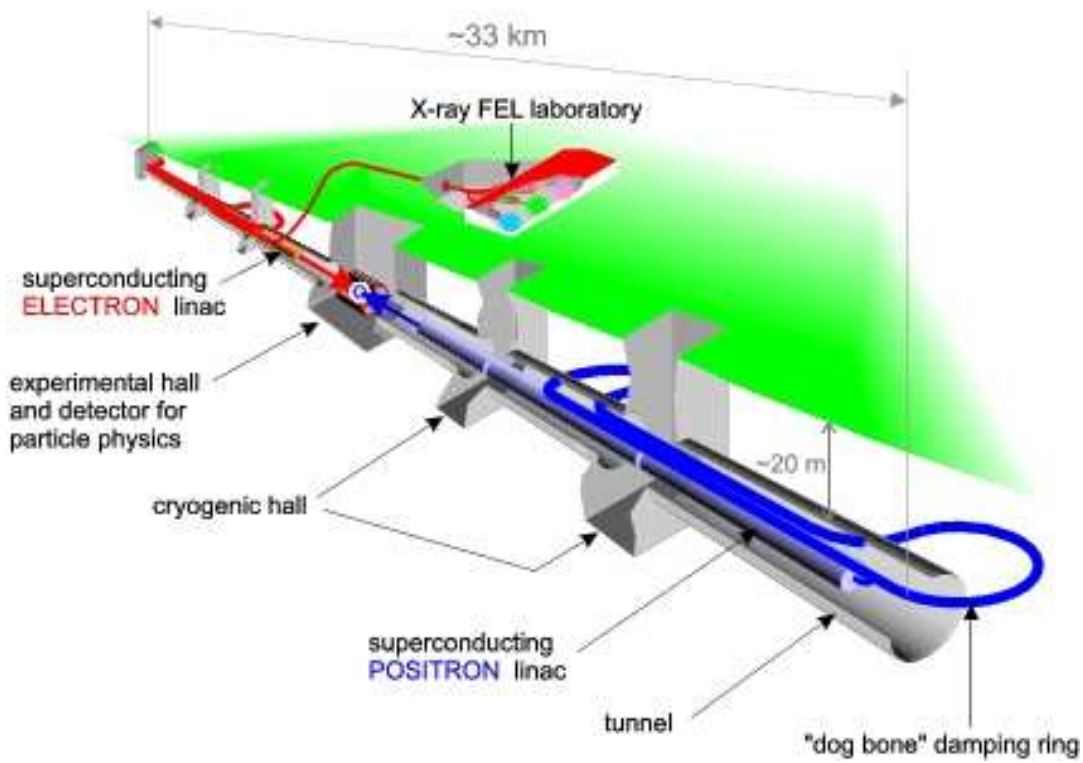}
\end{center}
  \caption{An artist's view of the TESLA electron--positron linear collider with integrated X-ray laser laboratory.}
\end{figure}
\addtocounter{exefigure}{1}
\addsec*{Revealing the Innermost Secrets of the Universe
}

Elementary particle physics has the ambitious goal to explain the
innermost building blocks of matter and the fundamental forces acting
between them. Symmetry principles determine the ordering of the
building blocks and the nature of their interactions. The masses of
particles and the exact strength of the forces played a key role in
the evolution of the universe from the Big Bang 
to its present appearance in terms of galaxies, stars, black holes, chemical
elements and biological systems. Discoveries in
particle physics thus go to the very core of our 
existence. 

Particle physics has made enormous progress in the past thirty years
by pushing back the frontiers of accelerators, experiments and theory.
Today we know that matter is composed of 
few fundamental building blocks, called quarks and leptons. A concise
theoretical framework for the forces between these constituents has
been developed which is based on the theoretical principle of gauge
invariance. By applying this principle it has become possible to unify 
the seemingly disparate 
electromagnetic and weak interactions into a single electroweak
interaction, and to develop a quantum field theory 
of the strong interaction, called Quantum Chromodynamics (QCD). The
forces are mediated by the photon, the $W$ and $Z$ bosons and the gluons,
the particles or quanta of the corresponding fields. 

However, such a theory of matter particles and force fields suffers
from a serious deficiency. The underlying gauge principle requires at
first sight all field quanta to be mass-less. This is in striking contrast to
the large masses of the $W$ and $Z$ bosons, which are 80--90 times
heavier than the proton. At present the only compelling way 
 to give the particles a mass while
preserving the gauge principle is the so--called Higgs
mechanism. The basic idea is that the a priori mass-less particles
acquire ``effective masses'' by interaction with a background medium,
the Higgs field. The idea of mass generation by the Higgs
mechanism leads to a Higgs field which spreads out in all space. 
An analogous mechanism is in fact responsible for 
the attenuation of a magnetic field
inside a superconductor. Associated with the Higgs field is a new 
observable particle, the Higgs particle, whose
analogue in superconductivity is the Cooper pair.

The matter particles, the force fields of the
electroweak theory and QCD, and the Higgs mechanism 
are the basis of the
so-called ``Standard Model'' of particle physics, 
an extremely successful theory which has been tested and validated with
high precision in a broad range of experiments. Only
the Higgs particle has so far escaped observation.

Therefore, one of the most pressing challenges of particle physics is
to establish the Higgs mechanism, to find the Higgs particle and 
to study its properties, or to find an alternative 
explanation of the masses of particles. 
With the help of the Heisenberg
uncertainty principle of quantum mechanics one can get a first glimpse
of the Higgs particle, even if it is too heavy to be produced directly. This
principle allows a heavy particle to appear as a ``virtual'' particle
for a tiny fraction of time and thereby to influence the
measurements. Precision measurements made at 
electron--positron colliders have confirmed this effect 
with two very important and striking results:
First, the mass of the top quark has been determined 
in experiments at colliders which did not have enough
energy to produce it directly, before it was observed in
proton--antiproton collisions at that mass.
This was convincing proof that the Standard
Model is correct even at the level of effects produced by virtual
particles, i.e., at the quantum level. Second,  
the upper limit for the mass of the Higgs
particle has been determined to be 
less than 200\GeV\ in the Standard Model. Recently, events observed
at the highest energy of 
the Large Electron Positron Collider (LEP) at
CERN in Geneva
have given a tantalising hint that the Higgs particle might indeed have a mass
around 115\GeV. In order to produce the Higgs particle directly 
and in particular, to study its properties precisely,
we require more energy, i.e. accelerators of higher 
energy than those available today. The Higgs particle is 
expected to play a central role in the experimental 
program at TESLA. 
%

Discovering the Higgs particle and establishing the Higgs mechanism 
would not, however,
close the book of particle physics because the Standard Model is based
on too
many assumptions and leaves too many facts unexplained. General
arguments clearly point to the existence of an even more fundamental
theory. Supersymmetry is the favoured idea underlying such an
extension of the Standard Model. It leads to a consistent and
calculable theory in which the Higgs
mechanism can be accommodated in a natural way. 
Most importantly, supersymmetry provides a framework for
the unification of the electromagnetic, weak and strong forces
at large energies. It is deeply related to gravity, the fourth of the
fundamental forces. Supersymmetry predicts 
several Higgs particles. The lightest Higgs particle
should have a mass below 200\,GeV, or even below 135\,GeV in some specific
models. In supersymmetric theories for every fundamental
particle we know today another related and as yet undiscovered
particle should exist. The lightest supersymmetric 
particle most likely is stable. Many of these supersymmetric particles 
may have masses such that they can be produced and studied in detail
at TESLA.

A particularly tantalising challenge of fundamental physics is posed
by gravity, the interaction responsible for the large--scale structure
of the universe. Gravity can not be incorporated 
in the Standard Model because gravity can not be formulated 
consistently as a quantum field theory.
Great efforts are devoted to
formulating a theory in which gravity is unified 
with the weak, electromagnetic and strong forces. 
A goal of such a fundamental theory will be to predict
the masses and properties of all particles
based on a few fundamental principles. It will 
synthesise  quantum physics and
the theory of relativity, thus unifying the physical laws of the 
microcosm and the macrocosm.
Recent theories, known as super--string theories,
suggest that at very high energies, as they existed shortly after the
Big Bang, all four forces between particles are united into a
single force. 
The discovery of supersymmetry and the precision measurements 
of the properties of supersymmetric particles could 
provide a glimpse of the underlying fundamental theory.

%

Probing matter at its smallest dimensions thus leads us to a better
understanding of the laws governing the cosmos. The theories of
particle physics describe matter also under extreme conditions, as
realised
during the earliest moments of the universe, when 
everything was very hot and dense.  Collisions of particles
in accelerators allow us to recreate in the laboratory what happened
immediately after the Big Bang, 15 billion years ago, when matter
in the form of quarks and leptons was
created from energy. Nature has
passed through this stage on the way to its current state. If we
succeed in determining these laws of nature 
we will get clearer insights into the current state of
development of the universe.

Astronomical evidence strongly suggests that more than 90\%
of the mass in the universe is invisible and of a nature totally
different from the matter from which stars, planets and humans are made of.
The nature of this so-called dark matter is completely unknown. 
Supersymmetric particles might be the explanation. In future large 
accelerators we expect to find such particles, if supersymmetry is 
indeed realised in nature.
Thus, in developing more powerful accelerators,
better experiments and theories, particle physicists contribute, together
with astronomers and astrophysicists, to the understanding of the
origin, evolution and destiny of the universe and the nature of space
and time.

The large accelerators in operation today are the electron--proton
collider HERA at DESY and the proton--antiproton collider Tevatron at
Fermilab near Chicago. The next milestone on the road of particle physics is set by
the large proton--proton collider (LHC), which is being built at CERN
 and which is scheduled to be completed in 2006. Many
new discoveries are expected to come from the experiments there.

However, our previous experience clearly shows 
that a proton collider alone is not sufficient to adequately  explore
the subatomic world. It must be complemented
with a high energy collider for electrons and positrons. A telling
example from the past is the heavy $Z$ boson, which
was discovered at a proton--antiproton collider, while its detailed
properties have only been measured with high precision at 
electron--positron colliders. These measurements were crucial for
establishing the Standard Model. In particular, they allowed an indirect
determination of the mass of the top quark and are responsible for our present
constraint on the Higgs mass. Another illustrative example is the
discovery of the carrier of the strong force, the gluon, which was not
seen at a proton collider, where the strong
force dominates, but was discovered 
at the electron--positron collider PETRA at DESY. 

The complementarity of proton and lepton colliders
is due to the different nature and properties of the particles which
collide in the two types of accelerator. Electrons and positrons have
no internal structure. Being fundamental particles 
they carry the full beam energy
and interact through weak and electromagnetic
forces, which can be calculated precisely. 
The conditions under which the collision takes place 
are defined very well, so that 
we can predict precisely what to expect after 
the collision. 
One can therefore 
determine the properties of new particles, such as mass, lifetime, spin and
quantum numbers, unambiguously and with high precision. On the other
hand, it is easier to accelerate protons to very high energies than it
is to accelerate electrons.
Protons are, however, complicated objects, composed of quarks, antiquarks and
gluons. 
The detailed collision process can not be well
controlled or selected, the effective energy of the colliding
fundamental particles is usually well 
below the total energy of the two protons,
and the rate of unwanted collision processes is very high. 
For these reasons TESLA complements the LHC and will provide important new
insights.

While most of the particle physics program will be using TESLA as an 
electron--positron collider, TESLA can also be operated to generate
photon--photon, photon--electron and electron--electron collisions
which would provide important additional insight. 
The electron beam of TESLA could also be used for other studies in particle and
nuclear physics, such as the analysis of the inner structure of the
nucleon and the properties of the strong force.

The electron--positron linear collider TESLA, in its baseline design,
reaches a centre--of--mass energy of 500 \GeV, five times higher than the
first linear collider SLC at Stanford and 2.5 times higher than LEP at
CERN. At the same time the luminosity of TESLA, a measure for the
event rate which a collider can deliver, is about 1000 times higher
than that of LEP at 200 GeV. Both, energy and luminosity are
prerequisites for new discoveries. In a second phase, the energy range
of TESLA can be extended to about 800 \GeV\  without increasing the
length of the machine. Parts~III, IV and VI of this Technical Design
Report present the detailed studies of the scientific case 
which have been performed and
which illustrate that TESLA will be a very powerful instrument to
substantially deepen our understanding of the microcosm and the
universe. 

The energy range and luminosity of TESLA will make possible precise
measurements of masses, lifetimes, and interactions of particles. These
measurements will be needed to understand the mechanism responsible for the
generation of mass. If the world is supersymmetric, with matter and
forces united in one theory, TESLA will be uniquely positioned 
to explore these new particles.
The great experimental precision
characteristic for electron--positron colliders can be exploited to
probe for physics in an energy range well beyond the reach of the
collider. The substantial understanding gained during the studies of
the physics potential of linear colliders 
provides us with a firm prediction: With TESLA
we expect unique and crucial new insights into the laws of particle
physics. As the technology to build TESLA
is available today it is now the time to start its
construction.

\addsec*{New Insights into the Facets of Nature and Life}

When the structural and electronic properties of matter are to be
studied on an atomic scale -- particularly when looking at atoms in
molecules, in large biomolecular complexes and in condensed matter --
then X--rays play a crucial role. Their wavelength is of the same
order as the inter-atomic distances, which makes them the ideal probe
for determining the structure on the atomic and molecular
scale. Furthermore, the penetration power of X--rays allows
determination of the true bulk behavior of matter. As a result X--rays
became one of the most important tools in basic science and medical
diagnostics, as well as in industrial research and development.

\begin{figure}[t!hb]
\begin{center}
\includegraphics*[height=10cm,bb=26 285 515 712]{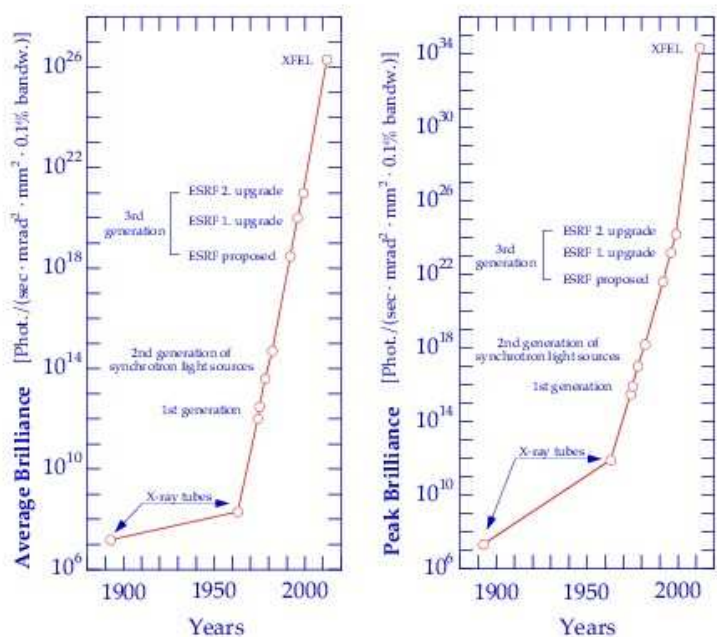}
\end{center}
\vspace*{-5mm}
\caption{\label{brilliance_in_time}
Increase of  time average and  peak brilliance of X--ray sources since
R{\"o}ntgen's discovery in November 1895. From the very beginning DESY has
been active in developing synchrotron radiation sources and, together
with external users, the appropriate instrumentation. (ESRF = European
Synchrotron Radiation Facility in Grenoble, 
XFEL = X--ray free--electron laser at TESLA).}
\end{figure}

Over the last thirty years tremendous progress has been made in X--ray
science and its applications, largely stimulated by the availability of
synchrotron radiation from storage ring facilities. Fig.~\ref{brilliance_in_time}
shows the gain in average and peak brilliance of the X--ray sources
over the last 100 years. So far, the dream of an X--ray laser in the
one {\AA}ngstr{\"o}m (0.1 nanometer) wavelength range has not yet become a reality. It is the
free--electron laser (FEL), which finally will provide lateral fully coherent
polarised X--rays with peak brilliances that are more than a 100
million times higher than what is available today from the best
synchrotron radiation sources. In addition, the X--rays will be
delivered in flashes with a duration of 100 femtoseconds or less,
allowing the observation of the fastest chemical processes. The
availability of lateral fully coherent, that is fully parallel, 
X--ray beams will also
stimulate the development of novel diffraction and imaging schemes. 

The generation of radiation in an FEL has much in common with 
the generation of radiation in a
conventional optical laser, the main difference being the gain
medium. In a conventional laser the gain comes from
stimulated emission from electrons bound to atoms, either in a
crystal, liquid dye or a gas, whereas the amplification medium of the
FEL is ``free'' (unbounded) electrons in bunches accelerated to
relativistic velocities with a characteristic longitudinal charge
density modulation. The radiation emitted by an FEL can be tuned over
a wide range of wavelengths, which is a very important advantage over
conventional lasers. 

The free--electron laser concept was first
realised for photons in the infrared wavelength range using a small
electron linac and a periodic magnet structure, called an undulator,
within an optical cavity. Later, in order to reach shorter
wavelengths, free--electron lasers were realised by installing such
optical cavities and undulators in electron storage rings. In this way a
wavelength of 190 nm has been achieved recently at the Elettra storage
ring in Italy. However, because of the strongly decreasing
reflectivity of the mirrors in the optical cavity one cannot expect to
reach wavelengths much below 150 nm with storage ring FELs. Instead,
in order to reach the one {\AA}ngstr{\"o}m wavelength range, a free--electron
laser concept without optical cavity is needed. Such a concept, the so
called Self Amplified Spontaneous Emission (SASE) scheme, has been
demonstrated down to wavelengths of 80 nm using electron bunches of
high charge density and low emittance from the linear accelerator at
the TESLA Test Facility. 

Linac driven X--ray FELs will open up
fundamentally new opportunities in science.
In his book ``From X--rays to Quarks'' Emilio Segr\`e writes: 
{\em In the 1920's we used to joke that good physicists, once passed
to their heavenly rewards, would find apparatus in paradise which,
with a twist of some knobs, would give electromagnetic radiation of
any desired frequency, intensity, polarisation, and direction of
propagation.} 
In a way he foresaw synchrotron radiation with its
enormous potential for studies in physics, chemistry and materials
science, for environmental and geo-sciences, as well as for structural
biology. With the X--ray FEL at TESLA there is an even more fascinating
vision and scientists address challenging questions such as:
\begin{itemize}
\item Can we take pictures of single macro-molecules?
\item Can we see the dynamical behaviour of the electrons as they form chemical bonds?
\item Can we make a movie of a chemical reaction?
\item Can we make real--time studies of the formation of condensed matter?
\item Can we make a movie of fast switching in magnetic storage devices?
\item Can we follow, for instance, a viral infection in a cell at high resolution?
\end{itemize}
In Part~V of the Technical Design Report these questions are discussed
together with a large number of further applications of the TESLA
X-ray free--electron laser.

Perhaps one of the most challenging, far--reaching applications
suggested for X-ray free--electron lasers is the imaging of nanometer
scale bio--molecular assemblies and the determination of their
structure with atomic resolution. The X--ray laser is expected to play
an important role in the structural and functional analysis of large
molecular complexes, which are crucial in the functioning of
cells, but which are extremely difficult to crystallise and can hardly
be studied with present day techniques.

Another example is condensed matter physics: The objectives of modern
condensed matter research are to determine the electronic
states of matter and its geometric structure on the atomic length
scale, to gain insight into the formation of condensed matter (either
indirectly via inelastic scattering or directly via real time
observations), to study the fundamental interactions in matter as well as
the relation between microscopic and macroscopic properties of
materials. This field of research forms the basis of modern materials
science and its engineering applications. For many questions related
to the study of novel materials, traditional techniques such as
neutron scattering or spectroscopy at present synchrotron light
sources face their limits of applicability, especially when trying to
understand ultra--fast processes on a nanometer length scale.

In short: Present day X--ray and neutron experiments probe in most
cases equilibrium states of matter. The X--rays from the free--electron
laser at TESLA probe the dynamic state of matter and will mainly be
used to study non--equilibrium states, and very fast transitions between
the different states of matter. These non--equilibrium states are of
eminent importance for many processes in biology as well as for
tailoring of materials properties in nano--scale devices.

\addsec*{Technological Breakthroughs as a Basis for New Research}


Accelerators have become a key tool to study the microcosm. Their
development started about eighty years ago, and has since been boosted
by many new ideas and technologies which extended the attainable
energy by a factor of around 10 every decade -- and thus the capacity to
resolve ever smaller objects and to create heavier particles. This
development of accelerators has led to important applications 
in other fields of science including medicine. Especially 
electron storage rings used for the production of X--ray beams 
of unprecedented brilliance provide key tools for modern research. 
Today they may be considered the most important 
spin--offs of particle physics driven accelerator research.
%

The most powerful accelerator concept to reach high energies is that
of colliders, in which particles are made to collide head--on. Among
existing facilities electron--positron colliders have played a very
special role. In the collision electrons and their anti--particles, the
positrons, annihilate each other and the resulting energy is converted
back into new particles, whose properties are measured in the
detectors surrounding the interaction point. Since electrons and
positrons are elementary point--like particles, they are the precision
tools of particle physics, providing accurate knowledge of the
reactions under study. These features were essential for many
discoveries made with electron--positron colliders, including new
quarks and leptons and the particle mediating the strong interaction,
the gluon, and for  precision tests of the Standard Model.

Except for the Stanford Linear Collider (SLC), electron--positron
colliders have so far been built as storage rings, the largest being
the Large Electron Positron collider (LEP)
at CERN, with a 27\,km circumference and a maximum energy per beam
of just over 100\,GeV. This concept, however, is not suitable for
reaching even higher energies, as electrons radiate electromagnetic
energy when forced on a circular path. The related energy loss
increases by a factor 16 when doubling the particle energy. 

Therefore the only way to reach electron energies substantially above
100\,GeV is by accelerating them on a straight line. This leads
directly to the concept of a linear collider, first proposed by
Maury Tigner in 1965. In this concept electrons and positrons  are
accelerated in opposite directions in two linear accelerators and made
to collide in the middle of a detector. Each linear accelerator
consists mainly of a large series of electromagnetic radiofrequency
resonators (cavities), which  efficiently generate the required electric fields
to accelerate the electrons and positrons.  
 
Over the past decades, several groups world--wide have been pursuing
different linear collider design concepts. Already in 1971 a group at
the Institute of Nuclear Physics in Novosibirsk started detailed
design work for a linear collider of several hundred GeV, addressing
many of the relevant problems. Several years later, groups at 
CERN,  at the Stanford Linear Accelerator Center (SLAC) in California,
and the Japanese National Laboratory for High Energy Physics (KEK)
in Tsukuba began work on linear collider designs. The feasibility of
the concept has been demonstrated by the successful operation of the
Stanford Linear Collider. All these concepts were based on
normal--conducting copper cavities.

A major challenge for all linear collider concepts is to obtain a
large collision rate (luminosity) of electrons and positrons at the
interaction point. This requires very small spot sizes of the beams at
the collision point and high beam powers.   

The TESLA approach differs from the other designs by the choice of
superconducting accelerating structures as its basic technology. As
will be shown, the TESLA linear collider based on superconducting
accelerating structures is ideally suited to meet the requirements
needed for a large collision rate, namely very small beam sizes and
high beam power.  The advantage of superconducting technology,
combined with the high efficiency to convert electrical energy to beam
energy,  has been acknowledged from the very beginning of the research
and development on linear colliders, but the technology was considered
to be considerably more expensive than conventional technologies. 

By a focused development program, started in 1992, the international TESLA 
collaboration in co--operation with industry succeeded in developing
superconducting microwave cavity structures which are capable of generating
an accelerating voltage per meter, called gradient, five times larger
than before 1990. In addition a reduction of the cost per meter of
accelerator by a factor of four was achieved. Together, these achievements
provide the basis for a realistic superconducting
linear collider with all its advantages. 

A prototype superconducting linear accelerator was built as part of
the TESLA Test Facility in order to gain long term operating
experience. To date it has been operated successfully for more than
8600 hours.

The development of a powerful linear accelerator for particle physics
has also created the ideal accelerator for a light source with
completely new properties: An X--ray free--electron Laser (XFEL)
producing X--rays with true laser properties, as first proposed for the
Stanford Linear Accelerator. The laser light is generated when
electrons travel through a special magnet structure, after having been
accelerated in a linear accelerator. The TESLA linear accelerator
based on superconducting cavities is ideally suited to provide
electron beams of the necessary quality. Using the TESLA Test Facility
accelerator laser light was generated for the very first time in the
wavelength range from 180 \,nm to 80 \,nm with an X--ray free--electron
Laser. This was a first proof of principle that such an X--ray laser can
be built and has stimulated intense activities in the field of XFELs
world-wide.

Summarising the work of the past decade the following milestones in
accelerator technology and development have been reached: 
\begin{itemize}
\item Cavities exceeding an accelerating gradient of 25\,MV/m are
being produced routinely by industry, thus fulfilling the needs for  a
500\,GeV collider.
\item Using a new surface treatment, gradients of greater than
40\,MV/m have been reached in single cell cavities, giving access to
energies of 800\,GeV.
\item The superconducting linear accelerator of the TESLA Test
Facility has been operated for more than 8600 hours.
\item The free--electron Laser principle has been demonstrated at
wavelengths of 80--180\,nm.
\item Other technologically challenging components needed for the
accelerator, like high--power klystrons, have successfully been
developed, built and operated at the TESLA Test Facility.
\end{itemize}
\noindent These successes provide the firm basis for this technical proposal.

\addsec*{From Vision to Reality}

TESLA opens new avenues of discoveries and addresses the most
important riddles, as formulated by Sir John Maddox, former editor of
Nature: {\sl What was the origin of the universe? What does matter
consist of? How did life originate and what is its nature?} 

The scientific potential of TESLA as an electron--positron collider and
an X--ray laser is far reaching and justifies the construction.

The TESLA project clearly demonstrates the close interaction and
inter--connection of different fields of science and technology. In
order to answer major questions of particle physics it became
necessary to push the technology of linear electron--positron colliders
considerably further, exceeding present facilities in energy and
luminosity. This development also paved the way for an X--ray free--electron
laser.
TESLA clearly illustrates how fields of science as far
apart as particle physics and biology can be advanced by a
breakthrough in accelerator development. TESLA will become the motor of
an innovative research centre.

The Deutsches Elektronen--Synchrotron DESY therefore proposes to the
international scientific community, to the German federal government
and to the northern German state governments (``L\"ander'') to build
TESLA in the vicinity of Hamburg.

Based on the experience gained in building the TESLA Test Facility and
on industrial studies the cost of the TESLA project for the baseline
design has been evaluated in detail for:
\begin{itemize}
\item the 500 GeV electron--positron collider: { \totalcost\  million \Euro} 
\item the accelerator components for the X--ray FEL: { \xfelacccost\  million \Euro}
\item the equipment cost for the undulators, beam lines and
experiments, including infrastructure -- 5 laser beam lines, each
equipped with 3 experiments, 5 other beam lines with 1 experiment each:
{\xfellabcost\  million \Euro}  
\item  one detector for particle physics:  \detcostone\  million
  \Euro\  to 
\detcosttwo\  million \Euro, depending on the choice of technology. 
\end{itemize}

The person--years required to build the accelerators amounts to \personyears,
and the construction time is estimated to be 8 years. 

Endeavours of the size and complexity of TESLA should be realised as
truly international projects. From its onset in 1992 TESLA was
therefore developed by a large international collaboration. The
intention is to build and operate TESLA, once it is approved, as an
international project of a limited duration of initially 25 years. A
possible model for the realisation of TESLA in an international
co--operation, a ``Global Accelerator Network'', is proposed as a basis for
further discussions.

\cleardoublepage

\addsec*{About the Authors}

The TESLA project has been developed in an international collaborative
effort and represents the result of eight years of work by many people.

The superconducting accelerator and X--ray free--electron laser (XFEL)
have been planned and developed by the TESLA collaboration, an
international collaboration of scientists, engineers and technicians
from more than 40 institutes in 9 countries. The collaboration has
jointly built the TESLA Test Facility TTF
at the Deutsches Elektronen--Synchrotron DESY in Hamburg, including a
linear accelerator, which has been operated for four years.

Scientists, engineers and technicians from the following institutes
have contributed either to the accelerator research and development or
to the construction and operation of TTF (listed in alphabetical order
of the country): 
 
\begin{center}
\begin{tabular}{lp{11cm}}
Armenia&       Yerevan Physics Institute\\

China &      IHEP Beijing, Tsinghua University\\

Finland&       Institute of Physics, Helsinki\\

France &       IN2P3/IPN Orsay, IN2P3/LAL Orsay, DSM/DAPNIA Saclay\\

Germany &      RWTH Aachen,
               BESSY Berlin, HMI Berlin, MBI Berlin, TU Berlin,
               TU Darmstadt, TU Dresden, Frankfurt University,
               GKSS Geesthacht,
               DESY Hamburg and Zeuthen, Hamburg University,
               FZK Karlsruhe, Rostock University, Wuppertal University\\

Italy&         INFN Frascati, INFN Legnaro, INFN Milano, INFN Roma 2 \\

Poland&        Inst. of Nuclear Physics Cracow, 
                Univ. of Mining \& Metallurgy Cracow,
                Soltan Inst. for Nuclear Studies Otwock--Swierk,
                Polish Acad. of Science Warsaw,
                Polish Atomic Energy Agency Warsaw,
                Warsaw University \\

Russia &       JINR Dubna,
                MEPhI Moscow,
                INP Novosibirsk,
                BINP Protvino,
                IHEP Protvino,
                INR  Troitsk\\

USA &           Argonne National Laboratory, 
                FNAL Batavia,
                Cornell University,  
                UCLA Los Angeles\\
\end{tabular}
\end{center}

The following institutes or organisations have contributed, together
with DESY, major hardware components to the TTF project:  Argonne
National Laboratory (USA), DSM/DAPNIA (France), FNAL (USA), INFN (Italy),
and IN2P3 (France). The development of the technology of
superconducting cavities was done in close collaboration with
CERN (Switzerland),
Thomas Jefferson National Accelerator Facility (USA), and KEK (Japan).

\newpage
The scientific case for TESLA and the related experiments have been
worked out by participants in a series of workshops on physics and
detectors for particle physics and on the many facets of applications
of X-ray lasers. The workshops in particle physics were organised
jointly by the European Committee for Future Accelerators (ECFA) and
DESY. The XFEL working group at DESY has organised 10 international
workshops on the scientific potential of the X--ray FEL.

In total 1134  authors from 304 institutes in 36 countries have
contributed to this Technical Design Report and its supporting
studies, without necessarily implying an institutional commitment.
The scientific community interested in TTF and TESLA 
continues to grow, as is 
the number of  collaborating institutes and countries.

\renewcommand\thefigure{\arabic{chapter}.\arabic{section}.\arabic{figure}}

\cleardoublepage
\addtolength{\parskip}{-3mm}
\tableofcontents
\addtolength{\parskip}{3mm}
\cleardoublepage
\pagenumbering{arabic}
\setcounter{page}{1}
\chapter{Particle Physics with the Electron-Positron Linear Collider}

The electron--positron collider TESLA in the baseline version reaches a
centre--of--mass energy of 500\GeV, five times higher than SLC, the
first linear collider built at SLAC, and 2.5 times higher than
LEP. The luminosity of TESLA, a measure for the event rate a collider
can deliver, is about 1000 times higher than that of LEP at 200\GeV. 
Both, energy and luminosity are prerequisites for new
discoveries. In a second phase, by adding more cooling and
radiofrequency power the energy range of TESLA can be extended to
about 800 \GeV\  without increasing the length of the machine. In
addition, with some modifications, TESLA can be operated 
with high luminosity 
at lower energies, between 90 and 200 \GeV \ centre--of--mass 
energy. 

The substantial knowledge of particle physics accumulated during the past
decades provides us with a firm prediction: We expect fundamentally
new discoveries in the energy range of TESLA. In this chapter we
summarise the present status of particle physics at the high energy frontier,
the open questions
and how our knowledge helps us define the next large step in particle physics.
We discuss the importance of TESLA for particle physics and cosmology.   

\section{What do we know today?}

For the last 30 years, particle physics has made dramatic progress in 
understanding the building blocks, the fundamental forces and the 
underlying symmetries of nature.
These achievements were based on intensive interaction
between experimental
discoveries and precision measurements at accelerators and the
development of a powerful theoretical concept: the electroweak theory
which gives a unified description of weak and electromagnetic
interactions and the theory of strong interaction, known as Quantum
Chromodynamics (QCD). 

The matter we are made of and which surrounds us is made of quarks and electrons. We have measured with great precision that three generations of quarks and leptons (electrons and neutrinos and their heavier partners) exist, which is of direct relevance for the Big Bang nucleosynthesis as the formation of light nuclei depends critically on the number of neutrino generations. 

The forces are described by so--called gauge theories based on a
fundamental symmetry principle, gauge invariance. The idea of
gauge invariance was introduced in classical electrodynamics. By
generalising the gauge principle it has been possible to show that the
apparently very different electromagnetic and weak forces are just
different manifestations of a single electroweak force which has 
been verified directly in electron proton scattering at HERA. This
unification of the weak and electromagnetic interactions is of the
same theoretical importance as was the merging of electricity and
magnetism into the theory of electrodynamics. The forces are mediated
by particles, the gauge bosons: photons, $W$-- and $Z$--bosons, and gluons.

Practically all of the many detailed and frequently very precise
experimental observations are described by a beautiful and consistent
theory, the {\sl Standard Model} of weak, electromagnetic and strong
interactions.

One of the most spectacular confirmations of the Standard Model was
achieved when the experiments at LEP and SLC succeeded in predicting
the mass of the top quark with remarkable precision although the
energy of the accelerators was too low for the direct production of this
heavy quark. Here use was made of a subtle quantum physical effect:
the Heisenberg uncertainty relation allows the production of the top
quark for a short instant of time as a so--called "virtual
particle". The effect of these virtual particles was measured in the
precision experiments at LEP and SLC.  
The prediction was verified by the subsequent
observation of  the top quark in proton--antiproton collisions at the
Tevatron. Establishing the validity of the Standard Model at such a
level of accuracy has been a major accomplishment and gives us
confidence in the predictive power of the theory. The
prediction demonstrated that the energy reach of an electron--positron
collider can exceed substantially its total energy.

In summary, particle physics is presently in an excellent, yet curious
state: on the one hand, practically all of the many detailed and
precise experimental observations are perfectly accounted for by the
Standard Model.
On the other hand there are serious gaps in our
understanding, like: 
\par

\noindent {\sl Which mechanism gives mass to the fundamental
particles?}\\[1mm]
\noindent Since Newton, mass plays a central role in physics, yet the physical
mechanism which generates this property of all matter has not been
established so far.
TESLA will allow clarification of this mechanism; in particular 
the generation of masses by the Higgs mechanism can be established
unequivocally.

\noindent {\sl Can the four fundamental forces of nature, the 
electromagnetic, the weak, the strong force and gravity, 
be unified in a comprehensive theory?} \\[1mm]
\noindent  A crucial step in finding the
answer to this question can be taken by embedding the Standard Model
into a supersymmetric theory in which matter and forces are united.
Supersymmetric theories have predicted the unification of
the electromagnetic, the weak and the strong forces at high
energies in excellent agreement with 
precision measurements. These unified theories also explain why the electric 
charges of electron and proton are identical to more
than twenty digits, vital for the stability of 
the matter which surrounds us. But they also predict that eventually
all visible matter is unstable. 
In the ever expanding universe it will turn
after a very, very long time
into photons,
electrons, neutrinos and their antiparticles.
Moreover, supersymmetric theories offer a rationale for the existence
of gravity, in the same way as gauge theories
do for electromagnetism. The high precision with which
the properties of novel supersymmetric particles can be
studied at TESLA, is indispensable for exploring this 
new particle world. 

Since the Standard Model gives an accurate description of all
experimental observations so far and at the same time leaves many 
questions unanswered, it provides a good starting point for 
extrapolations into areas where new physical phenomena might 
be expected. The fact that we can ask precise questions reflects the
high level of understanding that has already 
been reached by particle physics.

We know that the answers lie hidden at high energies. Part
of the answers will be found at existing accelerators and the 
large hadron collider (LHC),
however, for many questions the energy and precision of the 
TESLA linear collider is indispensable to progress in our understanding.

The physics program for electron--positron linear colliders in
the TeV range has been developed in the last decade through
numerous theoretical analyses and experimentally oriented feasibility
studies, many of which are presented in Part III of this Technical
Design Report. In the following we summarise a few of the central 
questions of particle physics and demonstrate how TESLA will contribute 
to answer them.

\section{The Origin of Mass}

The clearest gap of all in our understanding is the present lack of
any direct evidence how a fundamental symmetry as the 
electroweak symmetry is broken,
and how the masses of quarks, leptons and the force mediating 
particles are generated. Without answers to this
question the model is only an "effective theory" which is
incomplete. There must be an underlying theory which will explain
its apparently arbitrary features, including particle masses.

In a gauge--symmetric theory all fundamental particles, at first sight,
should be mass--less. Why are weak interactions then mediated by 
the very heavy $W$ and $Z$ gauge bosons? At present only one compelling
way is known to give the particles mass while
preserving the gauge principle: the so--called Higgs
mechanism. The basic idea is that the a priori mass--less particles
acquire an "effective mass" by interaction with a background medium,
the Higgs field. This idea leads to a definite and 
striking prediction: there should be a new particle, of 
a completely new kind, the Higgs particle.

\begin{figure}[t!]
\begin{center}
\includegraphics[height=6.5cm]{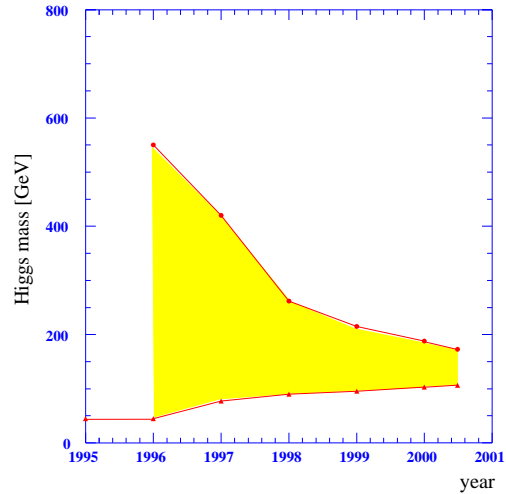}
\caption{\label{higgs_limits}
The allowed mass region for the Higgs particle (the
yellow region in the plot indicates the upper and lower 
$95\%$ confidence limits) has shrunk
considerably during recent years, due to the precision measurements at
LEP, SLC and Tevatron. The analysis is done in the framework of the
Standard Model.}
\end{center}
\end{figure}

Due to its central role the Higgs particle has been intensely
searched for at LEP and the Tevatron, but so far without success,
though tantalising hints have been reported at LEP. From existing
precision measurements and the quantum physical effects
mentioned above we can infer today, within the Standard Model, that
the Higgs particle should be lighter than 200\GeV\  (see
Fig.~\ref{higgs_limits}). 

\begin{figure}[tbh]
  \begin{center}
      \includegraphics[height=6cm]{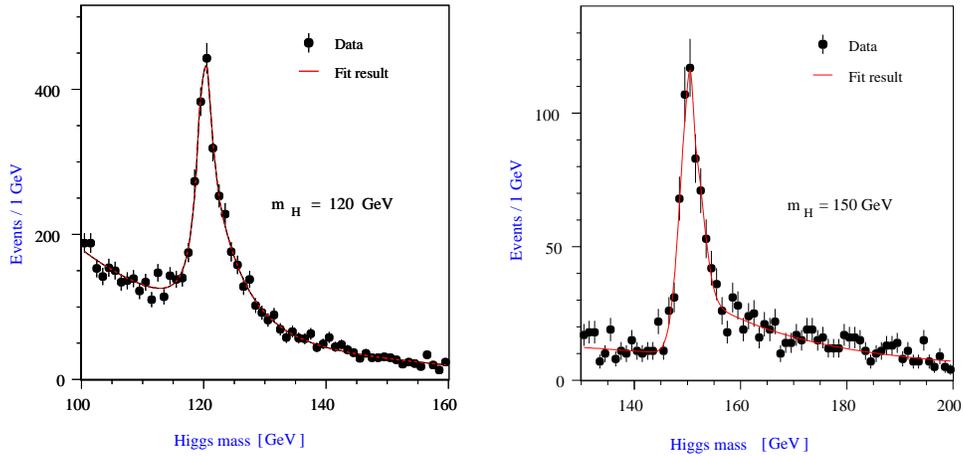} 
  \end{center}
\caption{\label{higgspeak}Signals expected for a Higgs 
particle for two different assumed Higgs masses. The 
distributions are nearly background free, illustrating the 
clean environment in which a Higgs particle can be 
reconstructed at the linear collider.}
\end{figure}

\begin{figure}[bt]
\begin{center}
\includegraphics[height=6.5cm]{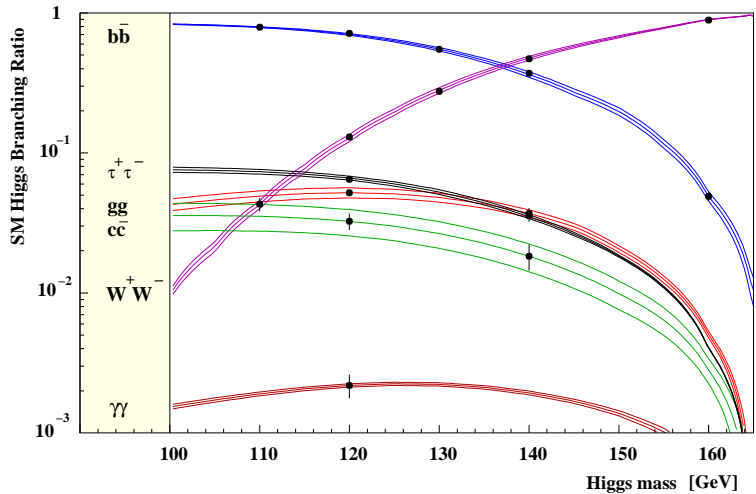}
\caption{\label{higgs_BR}
The predicted branching ratios of the Higgs particle in the Standard
Model (SM)
(i.e.\,the probability for the Higgs particle to decay into different
particles), as a function of the mass of the 
Higgs particle. The points with error bars show the expected experimental
accuracy which can be obtained after two years of data taking, 
while the lines show
the theoretical values and uncertainties of the Standard Model predictions.}
\end{center}
\end{figure}

Most probably the Higgs particle will be discovered at
the Tevatron or at the LHC. The precise
measurements of its properties however, indispensable for a complete
understanding of the mechanism by which 
masses are generated, require
a lepton collider.
With TESLA all the properties of the 
Higgs particle (see Fig.~\ref{higgspeak}) can be measured with high
precision: the mass, the lifetime, the production cross sections, the
branching ratios to quarks of different flavours, to leptons and to
bosons (see Fig.~\ref{higgs_BR}), and the way it couples to the top
quark. Moreover TESLA is unique in establishing the coupling 
of the Higgs particle to itself
which induces the electroweak symmetry breaking. 
A comparison with the properties
predicted by the Standard Model will establish whether or not
the Higgs mechanism
is responsible for electroweak symmetry breaking and test the self
consistency of the picture. TESLA will achieve a precision of
50\,(70)\MeV\  on the mass of a 120\,(200)\GeV\  Higgs, and will
measure many of the branching ratios to a few percent accuracy. The
Higgs coupling to the top quark will be measured to 5\%. 
The accuracy of all these measurements is vital for establishing the full
understanding of the origin of mass.

The option to add a second interaction region for photon collisions
will supplement the picture by precise measurements of the way the
Higgs particle couples to photons, which is a process particularly 
sensitive to effects from new particles with masses beyond 
the direct reach of TESLA.

If the Higgs particle does have the properties as predicted by the Standard
Model, the next stage of the program will be to refine even further
the existing precision measurements which constrain the model at the
level of effects introduced by virtual particles, i.e. at the 
level of quantum physical effects, 
by measuring the relevant parameters as precisely as
possible. One parameter is the  
mass of the top quark which can be measured at TESLA to an
accuracy of about 100\MeV. Other important constraints will come from
the value of the electroweak parameter $\sin^2 \theta_W$ and
from the mass of the $W$~boson. 
Both can be measured very precisely by lowering TESLA's energy
into the range between 90 and 200\GeV. 
TESLA can deliver a hundred times more data at these 
energies than LEP with the corresponding increase in accuracy. 
Inconsistencies between the Higgs properties and the parameters derived
from precision measurements of the electroweak bosons would give
direct information about physics scenarios beyond the Standard Model
at high energies.

\section{Supersymmetry: The Way to Grand Unification}

While theoretical arguments clearly point to the existence of a more
fundamental theory which incorporates the Higgs particle, at present we
cannot predict the energy needed to fully explore this underlying
theory. It may reveal itself entirely or in part at the LHC and TESLA or
may only appear through deviations observed between Standard Model
predictions and precision measurements done at TESLA. 

Supersymmetry (SUSY) is the favoured candidate for an extension of the
Standard Model because it preserves the successes of the 
Standard Model and provides a consistent and calculable theory
which can solve important theoretical issues. It eliminates the
problem encountered in calculating the quantum physical corrections to
the Higgs
mass and gives a natural explanation of the Higgs mechanism
responsible for the generation of masses. 

Most importantly, SUSY provides a consistent framework for the unification of
the three forces (electromagnetic, weak and strong) at very high
energy. When embedded in such a grand--unified theory,
the size of the electroweak parameter $\sin^2 \theta_W$ can be 
predicted very precisely. 
Its value has been confirmed experimentally at LEP at the per--mille level.  
Last but not least, supersymmetry is deeply related to gravity, the fourth of 
the fundamental forces.

SUSY predicts that each matter and force particle has a supersymmetric
partner, which has the same properties except the spin.
Each particle with integer spin has a
partner with half--integer spin and vice versa. As we have so far not
found any direct evidence for SUSY particles, their masses are expected to be
very large. The lightest SUSY particle may be stable. 

In contrast to the Standard Model, supersymmetric models include more
than one Higgs particle. The
lightest Higgs particle mass is predicted to be below 200\GeV, or even below
135\GeV\  in specific models. Measuring the properties of this particle
will reveal its origin in a new world of matter, the supersymmetric
world, and will shed light on the other heavy particles in the Higgs
spectrum which may lie outside the range covered directly by TESLA
(and the LHC). The experiments at the 
LHC can observe the lightest Higgs particle and access the spectrum of 
the heavy Higgs particles in certain circumstances. 
With TESLA
the Higgs particles can be directly observed if their masses are
below its highest beam energy (400 \GeV), or in photon--photon
collisions even beyond this limit. 
Within specific supersymmetric models TESLA's sensitivity 
can be extended to about 1\TeV\ through a precise measurement of 
the decay properties of the light Higgs particle.

\begin{figure}
\begin{center}
\includegraphics[height=7cm]{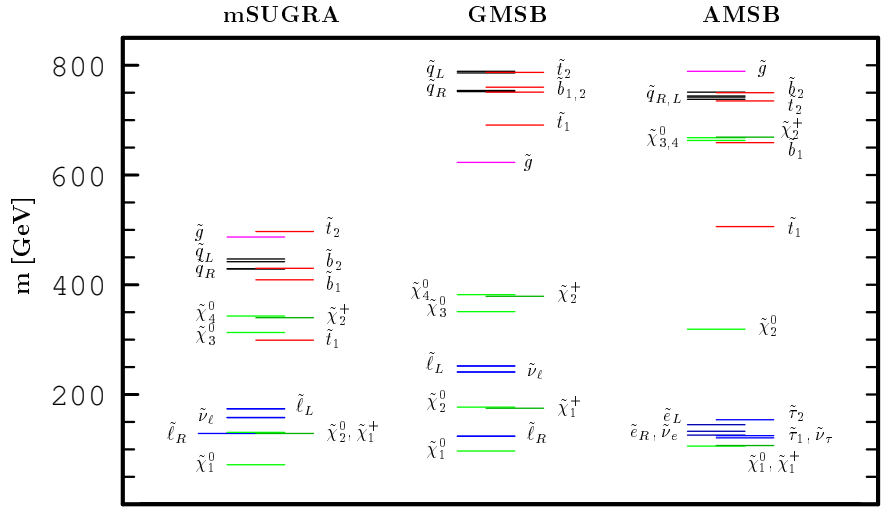}
\caption{\label{SUSY_spectrum}
Examples for the masses of supersymmetric partners of
leptons, quarks, gauge and Higgs particles, in three different
models. The abbreviations on top label these
models (mSUGRA: minimal supergravity, GMSB: gauge mediated
supersymmetry breaking, AMSB: anomaly mediated supersymmetry breaking).
Each line represents one particle, with its 
name indicated next to it, and the vertical scale
indicates the respective masses.}
\end{center}
\end{figure}

If supersymmetry is realised in nature an explanation is 
needed why it is not observed at low energies.
Several alternative theories have been developed which  
lead to a potentially rich spectrum of
supersymmetric particles within the reach of TESLA. Most of
the schemes predict light gauginos (these are the supersymmetric
partners of the photon, the $W$, the $Z$ and the Higgs particle) which TESLA should
be able to measure with high precision already in its
baseline configuration of 500\,GeV. As in optical spectroscopy, even
the observation of only parts of the spectrum will be sufficient to
establish unambiguously which SUSY model is realised in nature. LHC
on the other hand has an excellent potential to study the
supersymmetric partners of quarks and gluons.
Figure~\ref{SUSY_spectrum} shows some examples of mass spectra in
three representative models. Many of the predicted masses lie in the
experimentally accessible mass range.

The great variety of TESLA's precision measurements is required to
accurately determine the parameters of the supersymmetric theory. The
polarisation of the electron beam, available at TESLA, is particularly
important for these analyses. By varying the well defined
centre--of--mass energy of TESLA 
across the thresholds for new particle
production it will be possible to identify the individual particles
one by one and to measure supersymmetric particle masses with
very high precision. At the LHC part of the supersymmetric particle
spectrum can be resolved. Many final states however are 
overlapping which will
complicate the reconstruction of some of the supersymmetric
particles. Therefore, only the combination of the results from 
TESLA and the LHC will provide a complete picture.

\begin{figure}[!t]
  \begin{center}
    \includegraphics[height=8cm]{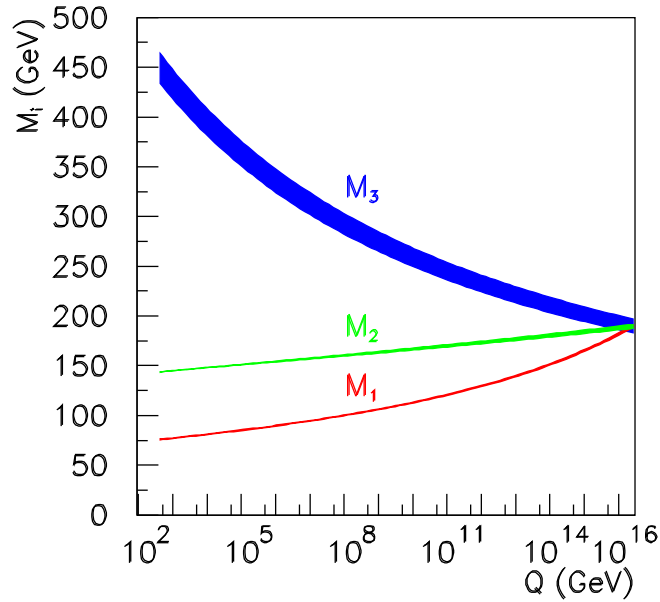}
  \end{center}
  \caption{\label{susy_gut}Evolution of the mass parameters for
  gauginos, the supersymmetric
partners of the photon, the $W$, $Z$ and Higgs particles, 
as a function of the energy scale Q, in a specific supersymmetric
  model (minimal supergravity). 
These mass parameters are 
proportional to the strength of the interaction.
At the unification
  scale all mass parameters should have the same value. 
 The results from the 
precision measurements done at TESLA energies ($Q\sim 10^3\GeV$) 
are extrapolated to high energies. The validity of the model can be 
tested by verifying whether the extrapolated mass parameters meet 
within errors at the unification scale. 
}
\end{figure}
The highest possible precision is needed to extrapolate the
supersymmetric parameters measured at the energy attainable with TESLA 
to higher energy scales, where the mechanism of supersymmetry breaking and the
structure of the grand--unified supersymmetric theory may be revealed
(see Fig.~\ref{susy_gut}). This is one of the most
important aspects of TESLA's physics potential, and may be the best way
to link particle physics with gravity through an experiment. 
\vspace*{2cm}

\section{The Link with Cosmology}

It has become increasingly clear that testing elementary particles
under extreme conditions in accelerators not only reveals the
basic building blocks and forces of nature,
but also sheds light on how the universe developed
during the earliest moments of its existence. Many aspects in the
research at TESLA are therefore of great relevance for both, particle
physics and cosmology. Two aspects have previously been mentioned in passing:
dark matter and the unification of forces. 

Understanding the unification of the weak, electromagnetic and strong
forces will tell us how during the process of expansion and cooling of
the universe one universal force has split into three distinct forces, 
each of which
plays a key role in our existence. 

Moreover, if the unification involved supersymmetry, as is generally
expected, one of the most intriguing scientific riddles may find a
natural explanation. Astronomy and astrophysics have revealed that
more than 90\% of all the mass in the universe must be of a kind entirely
different from the quark--based matter that makes up the stars, the
planets and ourselves. It neither emits nor absorbs light and
therefore cannot be seen in telescopes; it has appropriately been named
{\em dark matter}. Its presence is revealed by the gravitational
attraction that it exerts on the matter of celestial bodies. It
appears very curious that the dominant kind of matter in our
universe is of a nature completely unknown today. 
Supersymmetry however accommodates such a kind of matter in a 
natural way: Some of the particles in the hot primordial gas of 
the early universe could
be the lightest supersymmetric particle, which is stable in many
supersymmetric models. These particles effectively de--couple from
the 'normal' matter and have survived as dark matter, creating
substantial gravitational effects due to their large number and mass. With an
accelerator like TESLA these particles can be produced and studied
if they indeed exist.

Astrophysical observations have revealed another remarkable fact: 
In many of its aspects the universe appears as if it 
originated from a Big Bang driven by an inflationary expansive 
behaviour of a  
scalar field. Until now scalar fields have never been 
observed however, they are purely hypothetical. It is obviously of 
great interest to know whether such fields can and do indeed exist in nature. 
The Higgs field would be a generic example of a scalar field. Hence 
establishing the Higgs mechanism, one of the foremost research domains 
of TESLA, would be of great potential relevance also for cosmology.

\section{Alternative New Physics}

In spite of the many elements supporting the above picture which
incorporates a fundamental Higgs field for mass generation
and which can be extrapolated to high energy scales near the Planck
energy (which is of order $10^{19}\GeV$), 
there is no direct experimental proof that this is correct.
In fact numerous alternative theoretical ideas have been
developed, of which two concepts and their consequences for
the TESLA experiments have been analysed at some detail.

Kaluza and Klein first proposed many years ago to unify gravity and
electromagnetism. In this attempt they were led to assume that nature
has more than the four dimensions (three for space and one for
time) we seem to live in. This concept of
extra spatial dimensions, which we do not see, re--emerged
in the attempt to unify gravity and the weak,
electromagnetic and strong forces. So far these extra dimensions were
considered within a theory of quantum gravity, called super--string
theory, and were expected to be curled up into invisibility at
10$^{-33}$cm, a scale at which gravity becomes as strong as the other
forces, but also a scale totally inaccessible to any conceivable
accelerator experiment.

Very recently however it has been realised that one can consistently
introduce even macroscopic extra dimensions, at the level of micrometer,
without being in conflict with any direct observation. 
In such models a new mass scale appears which could be of order of only a few
\TeV. This has opened the possibility that we may be reaching a
new landmark in our quest for the fundamental theory.

The question naturally arises of how to test these ideas 
with accelerators, with neutrino beams, with micro--gravity experiments
or in cosmology. 
If this mass scale is indeed a few TeV, one can hope to observe various
signals directly both at TESLA and at the LHC. The observations at
TESLA will allow us to 
draw unambiguous conclusions.
Thus with TESLA we can tackle fundamental problems of the
structure of space and time. 

Although the Higgs mechanism in the Standard Model or its
supersymmetric extension remains the most compelling approach for the
generation of mass, there exist alternative schemes in which the
electroweak symmetry breaking is induced by new strong
interactions. Composite particles built up by new quarks, in the same
way as pions are made of quarks, would replace the Higgs field
and play a dynamical role in generating the 
masses of the electroweak gauge bosons. 
In this approach the interaction
between $W$ bosons becomes strong at energies close to one \TeV.
This would lead to anomalous values of the strength of the 
coupling between the
electroweak bosons, from which effective scales for the new strong
interactions can be extracted. Precision measurements of $e^+ e^-$
annihilation into $WW$ pairs at 500\,\GeV\  and of $WW$ scattering with TESLA's
high luminosity at 500 and 800\GeV\  have been shown to be sensitive to 
the onset of these strong interactions in a range up to 
$\sim$3\TeV. However theoretical scenarios of this kind are difficult
to reconcile with existing data
and they must be given rather complex
structures.
\vspace*{2.5cm}

\section{Challenging the Standard Model}

The importance of the precision measurements made at 
LEP/SLC/HERA/Tevatron has
already been mentioned repeatedly.  They provide a firm foundation
for our present understanding of the Standard Model. Yet within the
TESLA program it would be possible to achieve even greater precision 
on some of the quantities measured at LEP/SLC, which will
constrain the possibilities for new physics even more tightly. TESLA
operating at LEP energies could deliver a hundred times more
luminosity than LEP, 
with polarised electrons and positrons, neither of which were
available at LEP.

The physics impact of running at lower energies depends strongly on
the results obtained at 500\,GeV. It may happen for instance that at TESLA
and at the LHC only a single light Higgs particle is observed, and that 
no sign of any new phenomena are found.
With the precise knowledge of the Higgs and top masses and 
their properties derived from  measurements at 
TESLA (these masses dominate the quantum physical corrections in the Standard
Model) one could then verify the consistency of the 
theory with high precision:
Are these measurements
fully consistent with each other, 
or do we see signs for instance from 
new and very heavy particles or from extra dimensions 
which influence the measurements in a way which 
can not be explained by the Standard Model? 
This information would allow us to look far beyond the direct 
energy reach of TESLA.

\section{Colliding Light with Light or Electrons with Electrons}

Although TESLA is primarily conceived as an $e^+ e^-$ collider, it can
easily be transformed into an $e^-e^-$ collider.
A second
interaction region, which is not part of the baseline design, can be
either used to operate TESLA 
as an $e^+ e^-$ collider, as a $\gamma\gamma$ or as a $\gamma e^-$ collider.
Each of these different modes opens new experimental possibilities, both for
exploring the Higgs mechanism or supersymmetric theories. 
Further refinements of work now being done on QCD at HERA and with LEP
data may be envisaged. 
The TESLA design allows to realise these additional options. 

\section{Other Research Options}

Electrons from TESLA can also be used to explore the 
electromagnetic and hadronic structure of the nucleon
and the photon, and the properties of the strong force. These options
are called THERA, TESLA--N and ELFE. 

THERA uses the polarised and/or unpolarised electrons at the full TESLA
beam energy (or even at twice, when both accelerator arms are used) and brings them into collision with the 920\,GeV protons of HERA. This will be by far the most powerful electron microscope for the study of the structure of the proton and the strong force. It
gives access to a new domain not yet explored by HERA
and will contribute to answering
the fundamental question, why quarks and gluons are not observed as free
particles but confined in hadrons. It also opens a window to physics
beyond the Standard Model with a particular sensitivity to exotic
particles like lepton--quark and lepton--gluon bound states and excited fermions.
  
TESLA--N uses the interactions of the 250 - 400\,GeV longitudinally
polarised electrons of TESLA with a solid state target. ELFE
would use
15 - 25\,GeV electrons from TESLA, store them in HERA and finally
extract them as a
quasi--continuous beam onto a polarised target. The main goal of both
experiments is the precision measurement of a number of so far completely
unknown structure functions of the nucleon, which will widen our
understanding of its detailed structure and provide unique precision tests
of the predictive power of Quantum Chromodynamics (QCD).

\section{Doing Experiments at TESLA}

An electron--positron collision is a very well defined process. This
explains the key role electron--positron colliders have
played in the past for the progress of particle physics. Most of these
advantages stem from the following three unique strengths:

\begin{itemize}
\item A well defined initial state. This means that one knows that the interaction originates from an $e^+ e^-$ annihilation at a precise
energy. In the case of TESLA one can in addition define the spin
alignment (the polarisation) of the initial particles, providing a
powerful discrimination on electroweak interactions which depend crucially on
this alignment. 

\item Comparable rates for standard physics and new physics. Higgs
production for example has a rate comparable to other processes with
the same topology. 

\item Very favourable environment for measurements. Backgrounds are
low.  Particles can be observed very close to the collision
point, allowing for excellent precision on the decay points of
particles with short lifetimes. The final states of most events can be
completely reconstructed.
\end{itemize}

\noindent In Parts~III and IV these features are explained in greater
depth. The detector described in this TDR will allow almost perfect
reconstruction of most topologies, even with high complexity. Specific
properties of the detector are directly aimed at specific aspects of
the physics. As example, the identification of the final states
produced by bottom and charm quarks and by tau leptons plays an essential role
in Higgs physics. Furthermore, 
the detector has been optimised for a precise measurement of the 
energy of jets of particles, which are the 
experimental signature of quarks.

\section{Conclusion}

The successful experiments of the past decades and intense
theoretical work have led us to
the firm prediction of
new discoveries at the next generation of colliders, both the LHC and
an electron--positron collider like TESLA. The favoured prediction is a
light Higgs particle (with a mass most probably below 200\GeV), 
likely within a
supersymmetric extension of the Standard Model. This prediction is
supported by the existing precision measurements and by the
requirement that all forces including even gravity should be unified. At LEP
a tantalising indication of a Higgs signal 
has been observed which could be confirmed at 
the Tevatron or the LHC.
It appears most probable though not guaranteed 
that a light Higgs particle will be found. In this case TESLA will be the
ideal machine to test thoroughly all of its aspects. 
However regardless of the scenario nature has chosen 
a large variety of detailed studies confirm that 
TESLA will lead to new discoveries and key results in particle physics.

The detailed strategy for the experiments will depend on the
interplay of the results from LHC and TESLA:
\begin{itemize}
\item If there is only one single light Higgs particle within the
range of LHC and TESLA, its properties must be measured as
precisely as possible. One must establish whether 
this Higgs particle comes alone, whether it 
is a supersymmetric
Higgs particle, 
or whether it is the first sign of something completely new. The
essential next steps then will be to accumulate 
data at 800\,GeV and also at low
energies, around 90\GeV,  
in order to measure all  Standard Model parameters as
accurately as possible, and to search for possible 
deviations from the expectations.

\item If there is a light Higgs particle and if supersymmetric particles
are found in the energy range of TESLA, then the 
experiments at TESLA, combined with results from the
LHC, will be for supersymmetry what optical spectroscopy was for
quantum mechanics: the establishment of precision data from which the
underlying theory can be developed.

\item If there is no light Higgs particle, then supersymmetry is not
the correct theory at low energies. In this case there must be new 
strong interactions at a scale of a few \TeV.
The effects of such a theory would show up at the LHC and at TESLA.
The precision measurements at TESLA would provide a clear 
picture of the onset of the new interactions. 
\end{itemize}
Even if all these scenarios are not realised in nature, TESLA 
with its high resolution power can cope with the unexpected. 

In a study of the scientific case for a 500\GeV\  electron--positron
linear collider the American Linear Collider Working Group came to
very similar conclusions. 

The sensitivity of experiments at
TESLA may be improved 
by operating the collider as an 
electron--electron, photon--photon, or electron--photon collider, 
providing additional valuable information, for example for the study of 
the Higgs particle. 

All analyses 
require a machine delivering very high luminosity
and a high quality detector.
The TDR shows that this can be achieved with the TESLA collider and detector.

Lepton colliders will continue to play a key role in
the progress of 
understanding of particle physics and nature. To explore the energy
range beyond LEP the technology of linear colliders has to be pushed
to new limits. TESLA with its initial energy of 500\GeV, high
luminosity, the option to perform measurements of the $Z$ boson decays of
unprecedented accuracy, and its planned extension to 800\GeV\  is an
ambitious, but technically already well founded and justified step into
the future.  Any further step in energy will rely on the lessons
learned at LHC and at TESLA.

\chapter{The X-Ray Free Electron Laser Laboratory}

The X-ray free-electron laser laboratory proposed as part of the TESLA
project is conceived as a multi-user facility following the experience
of existing large synchrotron radiation facilities like the Hamburger
Synchrotronstrahlungslabor HASYLAB at DESY and the European
Synchrotron Radiation Facility ESRF in Grenoble.

At storage ring based synchrotron radiation sources a gain in
brilliance by more than ten orders of magnitude has been achieved over
the last thirty years. Each new generation of synchrotron radiation
facilities opened new, often fundamentally new applications, without
making the earlier applications less valuable. Therefore the user
community has been growing steadily and new storage ring based
facilities are under construction all over the world. World-wide the
number of users of synchrotron radiation is estimated at about 20\,000
scientists from many different disciplines: from physics, chemistry
and materials science, to structural biology and environmental and
geo-sciences. About 15 modern synchrotron radiation facilities
world-wide serve large multi-disciplinary user communities. Quick and
reliable access to a suitable experimental station is often decisive
for the success of the individual research project. The synchrotron
radiation community has therefore developed its own culture different
from that in particle physics.

Modern third-generation synchrotron radiation facilities provide
micro-focus beams of very high brilliance, with cross sections in the
sub-micrometer range, beams of almost complete circular or linear
polarisation, and beams with a significant degree of coherence. These
X-ray beam properties allowed to develop novel imaging techniques of
static and dynamic features of condensed matter including magnetic
properties, as well as the development of instrumentation for high
resolution inelastic scattering experiments to study phonon driven
processes or elementary electronic excitations. In protein
crystallography the structure of large macro-molecular complexes can
be determined, even in cases where only very small crystals are
available, and first sub-nanosecond time resolved studies have been
performed. After identification of the most promising research areas
the third-generation facilities put a lot of effort into improving the
quality of photon beams and optics, of instrumentation in general and
detectors in particular, in order to best serve the new demands of the
users. Today and for many years to come, research and development
activities at synchrotron radiation facilities are and will be
flourishing.
 
The storage ring technology itself approaches its theoretical limits
of performance with respect to average and peak brilliance, as well as
to minimal pulse duration. Making use of the principle of Self
Amplified Spontaneous Emission (SASE) the extremely high quality
electron beams of TESLA together with carefully designed magnet
structures (undulators) allow the production of laser-like X-ray beams
with wavelengths in the one {\AA}ngstr{\"o}m regime. Compared to present day
synchrotron radiation sources their peak brilliance is more than a 100
million times higher, the radiation has full transverse coherence and
the pulse length is reduced from the 100 picosecond down to the 100
femtosecond time domain. Thus, the X-ray free-electron lasers will
provide radiation of the proper wavelength and the proper time
structure, so that materials and the changes of their properties can
be portrayed at atomic resolution in four dimensions, not only in
space but also in time.

These outstanding research opportunities offered by linear accelerator
driven free-electron X-ray lasers create a deep excitement among a
rapidly increasing number of scientists all around the world. Many of
the early ideas developed in Italy, the USA and the former Sowjet
Union have been discussed in a series of workshops in the years from
1990 to 1994 at Sag Harbor, New York, and SLAC, Stanford, USA. In
February last year in a proof of principle experiment for SASE at the
TESLA Test Facility in Hamburg, lasing has been demonstrated for the
first time for wavelengths between 80 and 180 nm. At the LEUTL
facility at the Advanced Photon Source APS in Argonne, USA, saturation
of SASE has been reached for the wavelengths of 530 and 390 nm in
October 2000. In the same month the scientific case for the proposed
Linac Coherent Light Source (LCLS) to be built at Stanford, USA, has
been presented to the Basic Energy Science Advisory Committee (BESAC)
of the Department of Energy (DOE). It was favourably received and a
positive decision for appropriate funding is expected, so that
construction could start in fiscal year 2003. In May of this year a
collaboration of KEK (Tsukuba) and SPring-8 (Himeiji) will start an
ambitious R\&D program for SASE free-electron X-ray lasers in
Japan. Recently in Italy a research and development
program for linac driven X-ray FELs of the order of 
100 million \Euro\ has been launched.
 
\begin{figure}[tbh]
  \begin{center}
    \includegraphics[height=12cm,angle=-90]{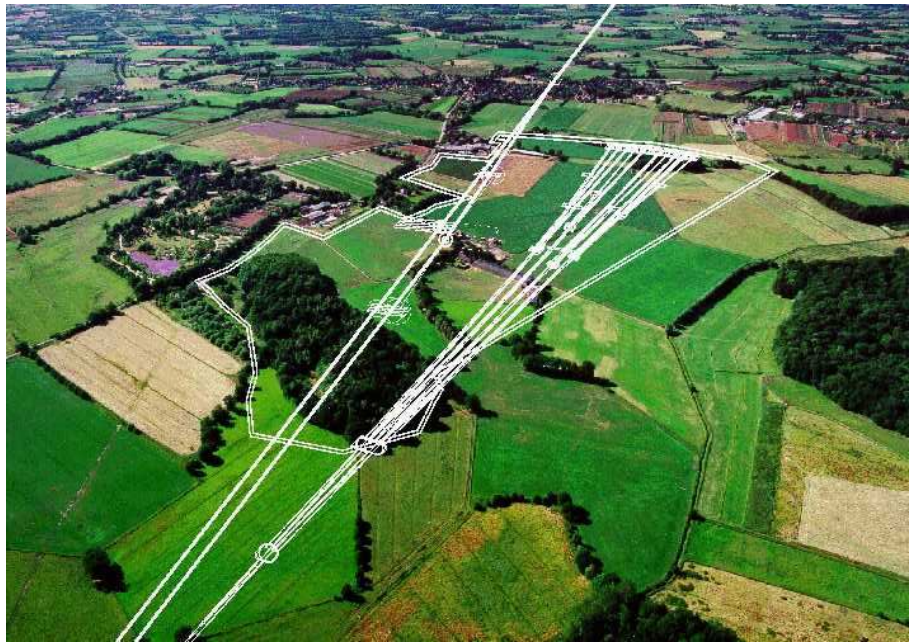}
  \end{center}
\caption{\label{fel_ellerhop}Aerial view of the research campus at 
Ellerhoop with an outline of 
the hall for the particle physics experiment and the 
X-ray free-electron laser laboratory superimposed. Most buildings 
are planned to be underground.}
\end{figure}

The recent successes at LEUTL and at the VUV-FEL at the Tesla Test
Facility, together with several other studies which have demonstrated
the validity of the SASE free-electron laser theory, give us the
confidence to propose an X-ray FEL laboratory with a large number of
experimental stations serving many users (Fig.~\ref{fel_ellerhop}). 
At present, the TESLA collaboration is in the forefront
of this exciting development of X-ray free-electron lasers. It is
expected that, as in the case of the third-generation X-ray
synchrotron radiation facilities like the ESRF in Europe, the APS in
the USA and Spring-8 in Japan, three mayor free-electron laser
facilities will be needed world-wide. Therefore, decisions for a
strategy towards a European multi-user FEL facility for hard X-rays
are urgently needed.
 
World-wide, a challenging R\&D program has been started in order to
learn how to make optimal usage of coherent FEL X-ray beams in the one
{\AA}ngstr{\"o}m wavelength range with highest brilliance and very short pulse
lengths. It is of great advantage that the FEL for vacuum ultra-violet
(VUV) and soft X-ray radiation at the TESLA Test Facilitiy at
DESY-Hamburg, reaching down to a wavelength of 6 nm in its
fundamental mode, will become available for users in 2004. In its third
harmonics this FEL will reach photon energies up to 600 eV. It will
permit a number of experiments of fundamental interest and will pave
the way for science at one {\AA}ngstr{\"o}m wavelengths at free-electron
lasers.
 
In the coming years we will also see strong synergy between R\&D for
X-ray FELs and laboratory based laser systems. For example, novel UV
laser systems have been developed for the photo cathode of the TESLA
X-ray FEL and for pump and probe experiments at the VUV FEL at the TESLA
Test Facility. Compared to modern storage ring based synchrotron
radiation facilities the average brilliance of laboratory based lasers
is still rather low and this difference is even rapidly increasing at
shorter wavelengths. However, in the UV and VUV wavelength range the
peak brilliance is becoming sufficient for preparing in university
institutes experiments in the femtosecond time domain, which will
finally be performed at appropriate wavelength and brilliance at a
facility like the X-ray free-electron laser laboratory at TESLA. The
interest in the science with X-ray free-electron lasers is growing
rapidly in the very large laser community. This community is familiar
with coherence, high power and femtosecond pulses of light at optical
wavelengths. Their fascination is due to the possibility to do similar
experiments at atomic resolution. In summary, the development of
linear accelerator driven X-ray lasers and the preparation of their
optimal use will play the role of a technology driver for various
disciplines in the years to come.

The proposed X-ray free-electron laser laboratory at TESLA will
consist of five FEL beam lines and five beam lines for spontaneous,
ultra-short pulse undulator radiation only. All together we expect
that up to thirty experimental stations can be distributed among the
ten beam lines according to the needs of the user community.

As discussed in Part~V of this report the TESLA X-ray laser will open
up most interesting new opportunities both for basic research and
applications in a wide variety of fields. In the following three
examples for research with an X-FEL are sketched: First, the study of
the early steps of formation and breaking of chemical bonds. Second,
the creation and analysis of new extreme states of matter. Third, the
new ways to determine structure and function of complex bio-molecular
assemblies.

\section{Analysis of Chemical Reactions at Ultra-fast Time Resolution}

Chemical compounds are described in terms of atoms, bond lengths and
angles. In order to describe a chemical reaction the molecular
structure and its evolution in time has to be known. This involves the
breaking and rearranging of intra- and intermolecular bonds for which
the time scale of fundamental steps is of the order of femtoseconds to
picoseconds and distances are typically in the {\AA}ngstr{\"o}m-range. Today,
extremely short wavelength lasers in the optical and ultra-violet
spectral range make it possible to observe dynamic events on the time
scale of a molecular vibration period. However, the lack of spatial
resolution on the atomic scale limits the entire field. Modern
spectroscopy can tell us how fast a molecular structure is changing,
but hardly how it is changing.

The importance of ultra-fast spectroscopy for studying chemical
reactions is increasing rapidly and has been recognised with awarding
the 1999 Nobel Prize in chemistry to Ahmed Zewail. However, since
optical spectroscopy only probes the energy of bound electrons, a
detailed theoretical knowledge of the energy levels of the system's
ground state and of all accessible excited states is required for
interpreting the spectroscopic data and indirectly deducing structural
information. Because of the large number of parameters describing
complex systems this knowledge can only be gained from these data by
computer simulations in the frame of accepted approximations. X-ray
based diffraction methods will provide new approaches for the
investigation of ultra-fast phenomena, complementing the information
accessible through ultra-fast optical spectroscopy alone. In
particular the combination of the two techniques is expected to give
for the first time detailed insight into the real-time formation of
chemical bonds.

In principle, table-top short-pulse X-ray and electron sources, when
utilised in pump-probe experiments together with visible
long-wavelength lasers, have the right wavelength to provide us with
such a view of chemical and physical transformation processes. 
However, their brilliance and temporal resolution is by far too
low. The TESLA X-ray FEL is ideally suited to the purpose of this
emerging scientific field. Its brilliance is many orders of magnitude
higher than achieved by table-top X-ray sources and the temporal
resolution of the order of 100 femtoseconds will make it possible to
resolve events on sub-vibrational time periods for many molecular
systems. Utilising a set-up where e.g. an ultra-fast optical laser
pulse initiates a photo-reaction which then is probed with the X-ray
laser pulse, enables one to follow ultra-fast structural changes
accompanied by electronic rearrangements, bond breakings and bond
formations. These are exactly the processes which determine some of
the most important chemical and biological reactions.

\begin{figure}[tb]
\begin{center}
\includegraphics[height=6cm,clip=true]{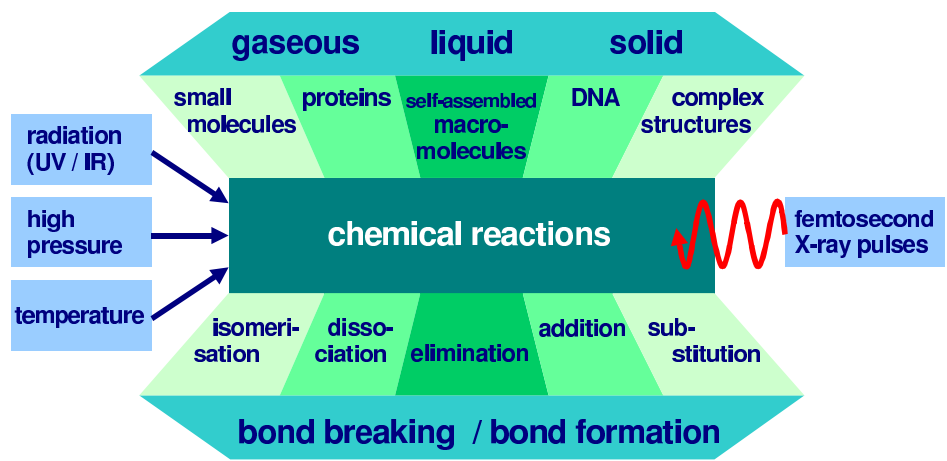}
\end{center}
\caption{Fields of applications for time-resolved investigation of chemical
reactions. Systems in different phase (top) can undergo various
chemical processes (bottom), which can be triggered by a variety
of methods (left) and investigated by femtosecond X-ray pulses
(right). \label{fel-pre:figt1}}
\end{figure}

Fig.~\ref{fel-pre:figt1} summarises possible experimental schemes and their
applications in the field of classical chemistry (organic, inorganic,
physical) and biochemistry, as well as materials science. After the
initiation of a chemical reaction its mechanism can be followed with
one X-ray pulse alone or a time sequence of X-ray pulses. Due to the
high flux provided by the X-ray laser, the time-resolved behaviour of
small molecules as well as of complex large molecules can be examined
in the gaseous, liquid and solid phases.

Today, light-triggered time-resolved studies with 100 picoseconds
resolution are performed at third generation synchrotron radiation
sources. However, recent results in ultra-fast optical spectroscopy
lead to the conclusion that the exact description of reaction dynamics
cannot be obtained within the usual theoretical approaches such as the
Born-Oppenheimer approximation which firmly links the nuclear and
electronic motion during the transformation. In particular in all
processes, where electrons are excited into delocalised states
(e.g. in semiconducting systems, aromatic chromophores like in the
light harvesting complex of the photo-reaction center) this picture
seems to break down. Especially here, more detailed studies about
structural rearrangements are needed.

\begin{figure}[tb]
\begin{center}
\includegraphics[height=9cm,clip=true]{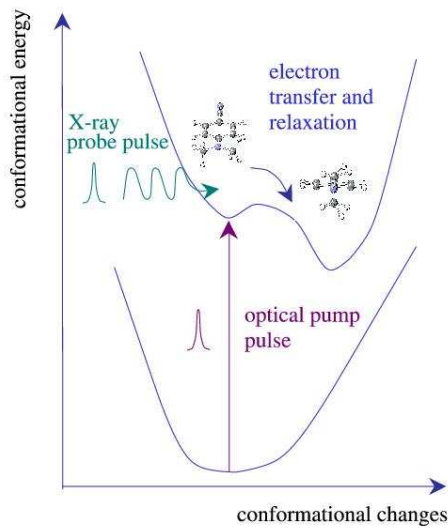}
\end{center}
\vspace*{-8mm}
\caption{Conformational rearrangements of a photo-excited
N,N-dimethylaminobenzonitrile molecule. After excitation with an
optical femtosecond laser as a pump the molecule changes its
conformation by electron transfer and relaxation processes. Probing
the excited state with X-ray FEL radiation without affecting the
decay process, allows to follow this conformational change in time.
\label{fel-pre:figt3}}
\end{figure}

A femtosecond light excitation may synchronise molecules in the sample
for a couple of picoseconds. This phenomenon offers amazing new
possibilities for structural investigations with femtosecond X-ray
pulses through the observation of coherent reaction dynamics. On
longer time scales chemical reactions do not proceed
synchronously. Here intermediates may be present simultaneously, but
they are vibrationally decoupled and thus unsynchronised. The
interpretation of all time-resolved experiments outside the
femtosecond domain is invariably compromised by this factor today. The
TESLA X-ray FEL will allow to study the structural reorganisation
processes related to chemical reactions in all phases, including
liquids. These experiments will profit both from the accessible time
range and the much increased number of X-ray photons per pulse, which
will yield far better photon statistics. This is particularly
important when considering ultra-fast phenomena, as the magnitude of
the structural motions yield signals, which are likely to be much
smaller than those occurring on slower
time-scales. Fig.~\ref{fel-pre:figt3} one
example of a photo-induced relaxation process is shown, which is the
rearrangement of a N,N-dimethylaminobenzonitrile molecule induced by
an electron transfer process. With the X-ray laser it will be possible
to follow such ultra-fast structural reorganisations, providing
information about the picosecond dynamics of this system at atomic
scale resolution.

A broad range of chemical reactions occurs on slower time scales. Most
of the reactions are not triggered by light but by temperature-jump,
pressure-jump or a combination of both. Currently the available X-ray
flux is not sufficient to follow non-cyclic reactions (i.e. reactions
which do not repeatedly go through the same cycle of processes) on the
nanosecond or even microsecond time-scale. With the X-ray FEL
time-resolved X-ray diffraction and absorption experiments can be
performed on non-cyclic reactions. This will enable chemical, thermal
and pressure triggered processes to be used for initiation of the
reaction, in particular for systems with complex structural response
characteristics like self-assembled macromolecules, proteins or
DNA. In combination with the flexible bunch-to-bunch time structure of
the TESLA X-ray FEL, there will be sufficient X-ray photons per pulse
to yield high quality structural information from a single
exposure. This approach will open windows for investigating a huge
domain of dynamic chemical phenomena, which cannot be structurally
characterised with current X-ray sources or classical spectroscopic
tools.

Experimental and theoretical studies of electronic relaxation
processes leading to molecular reactions and/or phase transitions in
condensed matter are among the most exciting goals in molecular and
condensed matter science. These processes occur on the femto- to
sub-femtosecond time scale and it has been demonstrated that optical
spectroscopy with ultra-short laser pulses can provide new information
on the nature of chemical reactions, however, atomic scale resolution
is missing. Novel experiments and theoretical approaches are needed to
gain better understanding of these extremely fast processes. The
excellent intrinsic time resolution in the sub-femtosecond
spike-structure of the X-ray pulses produced by SASE free-electron
lasers, as well as the application of a time-slicing techniques to
reduce the pulse duration to 10 femtoseconds or even shorter, hold
high promise for the detection of space-time fluctuations of the
electron density within chemical bonds. Such changes of the electronic
structure of specific elements at specific sites could be measured by
means of X-ray spectroscopy on the appropriate time scale.

\section{A Tool for Plasma Physics}

\begin{figure}[thb]
  \begin{center}
    \includegraphics[height=8cm]{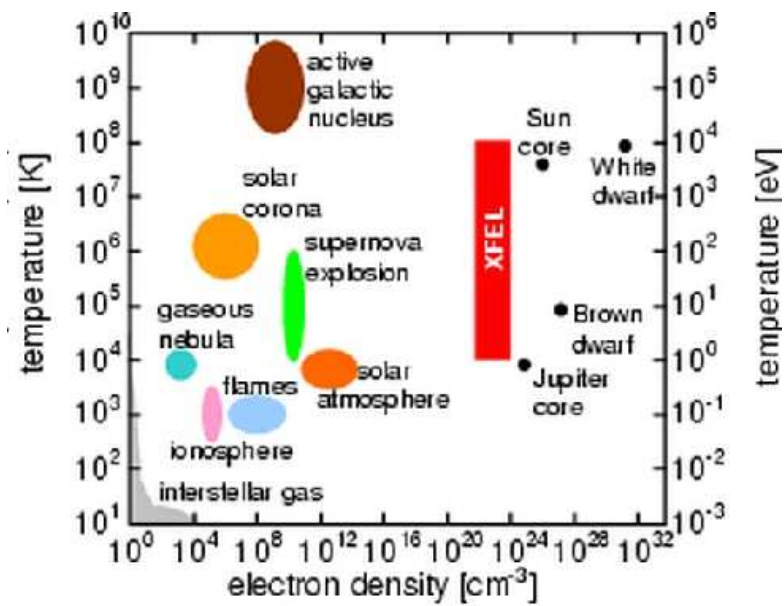}
  \end{center}
  \caption{\label{exesum_cosmos}Density--temperature diagram for
    astro--physical objects and the range accessible by the TESLA 
X--ray free--electron laser.}

\end{figure}

The X-ray laser, with its high brilliance and high energy density per
pulse, is also a powerful instrument to generate and probe extreme
states of matter in form of strongly coupled plasmas. It will allow to
heat much bigger volumes uniformly than those possible by other means
to unprecedented states of plasma with temperatures up to the keV
level ($10^7$\,K) and pressures up to the Gbar ($10^{14}$\,Pa) range. In a more
general sense the properties of these plasmas are closely related to
states of matter like in the interior of Brown Dwarfs, main sequence
stars, White Dwarfs, or the giant planets such as Jupiter or the
recently found extra-solar planets. Understanding the behaviour of
matter under extreme conditions is of fundamental importance to
develop improved models for planetary and stellar structure and its
formation, as well as for understanding the evolution of the universe
(see Fig.~\ref{exesum_cosmos}).

Because hydrogen and helium (and their mixtures) are the simplest and
most abundant elements in nature it is of special interest to study
them in the plasma state over a wide range of densities and
temperatures. For instance, metallic like conductivities in hydrogen
have been observed for the first time in dense hydrogen fluid at
pressures around 1.4\,Mbar and temperatures around 2500\,K in multiple
shock-compression experiments. Furthermore, substantial deviations
from standard equations-of-state have been found. Such data provide in
many cases the basic ingredients for inertial confinement fusion,
which employs both laser driven or heavy ion driven compression of
deuterium-tritium capsules. The X-ray FEL will provide new diagnostic
methods to determine the equation-of state independently of model
assumptions.

Today dense, strongly coupled plasmas can be produced by using
heavy-ion beams, as well as by high-power or ultra-short pulse
lasers. For instance, optical laser pulses of 100-1000 femtoseconds
duration and about 200-1000\,nm wavelength generate plasmas by photo-
and field-ionisation of valence electrons of atoms in a thin surface
layer. The X-ray free-electron laser laboratory at TESLA will provide
beams of 100 femtosecond pulse duration and with much shorter
wavelengths (0.1-2.5\,nm). In contrast to optical wavelengths radiation
the X-ray pulse will interact with the inner shell electrons of atoms
and the pulse will penetrate the entire sample. The power can reach
$10^{18} - 10^{20}$\,W/cm$^2$. The X-ray FEL will allow to study specific plasma
states for a variety of parameters and thus provide the data needed
for comparison with the results of computer simulations studying the
generation and evolution of plasmas including radiation transport and
hydrodynamic expansion. The data will also allow for a better
theoretical description of the interaction between energetic X-rays
and cold, warm, and hot matter. Focused X-ray FEL pulses, which will
be available at the TESLA X-ray FEL laboratory, will convert solid
target matter into a dense plasma at record temperature / pressure
values and, thus, open a new branch in dense plasma research.
 
In summary, the X-ray free-electron laser laboratory at TESLA opens a
new path to generate and diagnose dense and strongly coupled plasmas
by hard X-rays, which penetrate into the sample and heat it. This is
due to the high energy density and the short pulse duration of the
radiation, which produce a plasma in the volume of the target with
minimum gradients rather than in a surface layer as produced by
optical lasers. Thus, the production of dense, strongly coupled
plasmas by means of the X-ray FEL combines the advantages of using
ultra-short laser pulses with those of using heavy-ion beams. 

\section{Opportunities for Life Sciences}

During the last years the face of biology has changed dramatically and
new knowledge is acquired at an astonishing rate. Structural analysis
on a molecular level has become an important tool to study the
functions of individual components of biological systems in
detail. This refers in particular to protein crystallography revealing
the three-dimensional structures of many proteins and even of large
multiple-protein assemblies at atomic resolution. Structural results
provide a basis for the development of new drugs and offer novel
therapeutical procedures. Atomic resolution insight has, however, been
limited to bio-molecules that can be crystallised.

These days the sequence of the human genome is almost fully
determined, the genomes of many other organisms are known already. The
next step is to interpret this information, to determine the proteins
encoded therein and to investigate their function. Structural and
functional genomics projects are trying to address these issues. It is
anticipated that the main applications of the X-ray FEL in this field
will be in the post-genomic era, where the extreme intensity and the
very short duration of the pulses could open up new avenues for
investigations, especially on non-repetitive and non-reproducible
samples, like living cells. The X-ray FEL will open new possibilities in the
study of biological systems and their function. However, this path of
accumulating more and more information is not free of obstacles.

For almost any functional analysis the three-dimensional structure of
a macromolecule is required. However, preparing crystals of sufficient
size and quality for diffraction is one of the main bottlenecks to
retrieve high resolution structure models. This is one of the main
reasons that three-dimensional protein structures can presently be
obtained from only 15-25\% of the cloned genes.

A special case are membrane proteins where only a small number of
three-dimensional structures are known today. They have hardly any
tendency to form crystals. Various estimates show that there are at
least as many different membrane proteins as soluble proteins, yet
there are only a handful of independent structures known today for
integral membrane proteins, simply because they cannot be
crystallised. However, an understanding of the structure-function
relationships in membrane proteins would make invaluable contributions
to biochemistry, physiology and medicine, and would produce a
substantial socio-economical impact (about 75\% of all known drugs act
on membrane proteins).

Another very interesting and important class of objects are multiple
protein, RNA, DNA complexes. From the structure and behaviour of these
complexes, information on the mutual interaction of the individual
components can be deduced and will show how biology works on a
molecular or atomic level. Like membrane proteins these large
complexes are often very difficult to crystallise. Therefore, any
method capable of retrieving high resolution structural information
without the need for conventional diffraction-size crystals would be
extremely helpful. 

For both cases X-rays from free-electron lasers have the potential
to provide such insights. The large number of photons per pulse, the
very short pulse width and the possibility to focus to extremely small
focal spots open the way for diffraction from nano-clusters of
macromolecules, nano-crystals, single virus particles, two-dimensional
arrays of macromolecules, or of large macromolecular complexes. Also
the analysis of a number of individual macromolecules attached to some
sort of scaffold of known structure may become feasible.

\begin{figure}[t!b]
\begin{center}
\includegraphics[width=13cm]{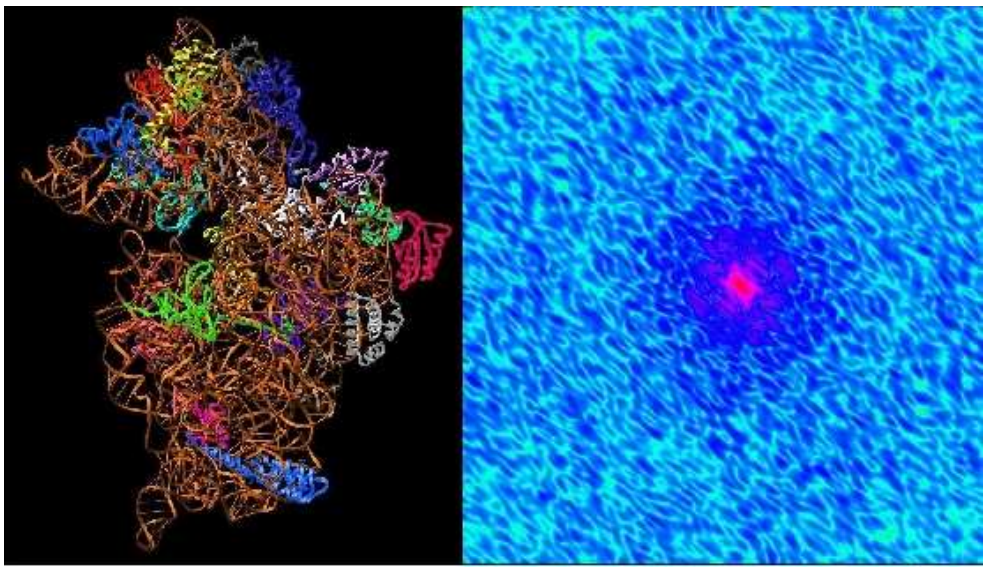}
\caption{ Typical example for the structure of a large macromolecular
  assembly and its molecular transform. Left: Structure of the small
  30S subunit of the Thermus Thermophilus ribosome; Right:
  slice through the 3-dimensional molecular transform of this
ribosome, which is closely related to the distribution of the 
scattering power in momentum space;
  this pictures corresponds to a spatial resolution of 
  around 4\AA.
} \label{ls-sum:fig4}
\end{center}
\end{figure}

The power density in the focused beam conditions would be extreme and
no sample will survive very long after the passage of a single pulse
due to ``Coulomb explosion'', since the incoming radiation will ionise a
large fraction of the atoms. However, as the pulse can be much shorter
than the initiation time of the Coulomb explosion it can in turn be
possible to record a diffraction image before the atoms leave their
original positions, and the sample is destroyed. Alignment procedures
similar to those developed in image reconstruction for cryo-electron
microscopy will help to obtain averages from a large number of
individual images and provide a high resolution molecular transform of
the molecule of interest (see Fig.~\ref{ls-sum:fig4}). The continuous molecular
transform, which is the modulus square of the Fourier transform of the
molecule, is significantly over-sampled as compared to the information
that can be measured from a crystal. In this case the phase problem
can be circumvented and the solution of the structure is
straightforward. Due to the coherence properties of XFEL radiation one
can also envisage holographic imaging techniques. 

\begin{figure}[!t]
\begin{center}
\includegraphics[width=9cm,angle=-90]{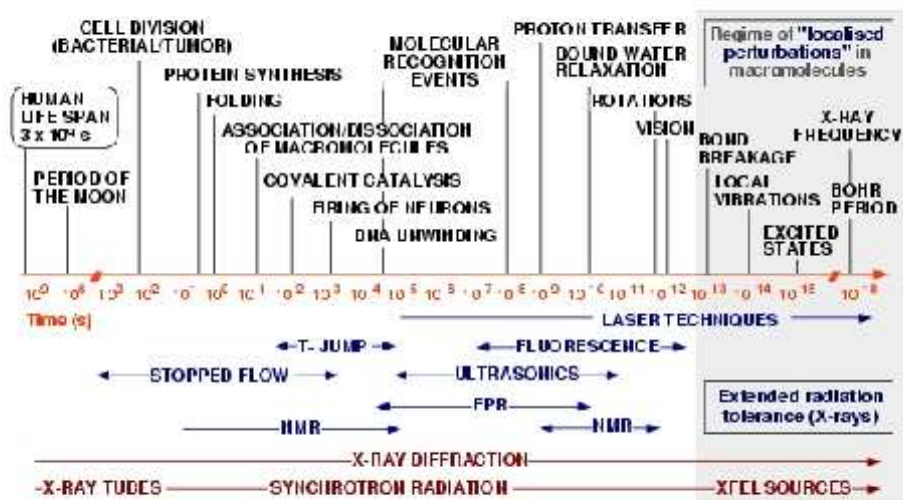}
\vspace*{-5mm}
\caption{ 
Relevant time scale for a number of biochemical processes. With
present day sources time scales down to $\sim 10^{-9}$ seconds are accessible
albeit not with enough single pulse intensity. This time-scale is much
too long to study key events at the heart of chemical
transformations. The fastest bio-chemical processes occur within a few
hundred femtoseconds. Structural experiments in this time domain will
be possible with the XFEL radiation, and will lead to a synthesis of
femtosecond spectroscopy and femtosecond structural methods. Such
studies cannot be done without an X-ray free-electron laser
laboratory.
\label{ls-sum:fig1}}
\end{center}
\end{figure}

Macromolecules can have a wealth of different functions, which in most
cases can not be derived from the static three-dimensional structure
alone. The typical time scale of biological processes spans many
orders of magnitude (Fig.~\ref{ls-sum:fig1}). With its very short, 100
femtoseconds radiation pulses and wavelengths down to one {\AA}ngstr{\"o}m the
X-ray FEL provides the potential for a time resolution three orders of
magnitude better than at any present day source. A femtosecond
excitation may synchronise molecules in the sample for a short time (a
couple of picoseconds). This phenomenon offers amazing new
possibilities for structural investigations with femtosecond X-ray
pulses through the observation of coherent reaction dynamics. 

Investigations concerning the dynamics of macromolecules exploiting
inelastic scattering effects (e.g. nuclear resonant inelastic
scattering) are nowadays significantly limited by the number of
photons available in the very narrow band pass required. The seeding
option for the X-ray FEL will provide an ideal tool for inelastic
studies with an intensity in an extremely narrow wavelength band,
which is orders of magnitude higher compared to third-generation light
sources.

In summary the X-ray free-electron laser laboratory at TESLA will give
us an entirely new tool permitting to determine the three-dimensional
structures of macromolecules without the need for diffraction-size
crystals, and to identify and characterise them in their cellular
environments. A number of technically challenging issues have to be
solved in order to fully exploit the potential of the X-ray
free-electron laser for biology. Basic aspects of the interaction of
femtosecond X-ray pulses of extreme power density with atoms and
molecules in soft matter, including non-linear effects, have to be
studied. Radiation damage and its evolution on the femtosecond time
scale will be investigated and many biochemical preparation techniques
have to be improved or novel ones have to be developed in order to
provide well defined samples without changing their physiological
properties. One may foresee that ultra-short, high intensity X-ray
pulses, in combination with novel container-free sample handling
methods, will open up new horizons for structural and functional
studies on both the molecular and cellular level.

\chapter {The Technical Layout}

\section{Introduction
}
\begin{figure}
\begin{center}
\epsfig{file=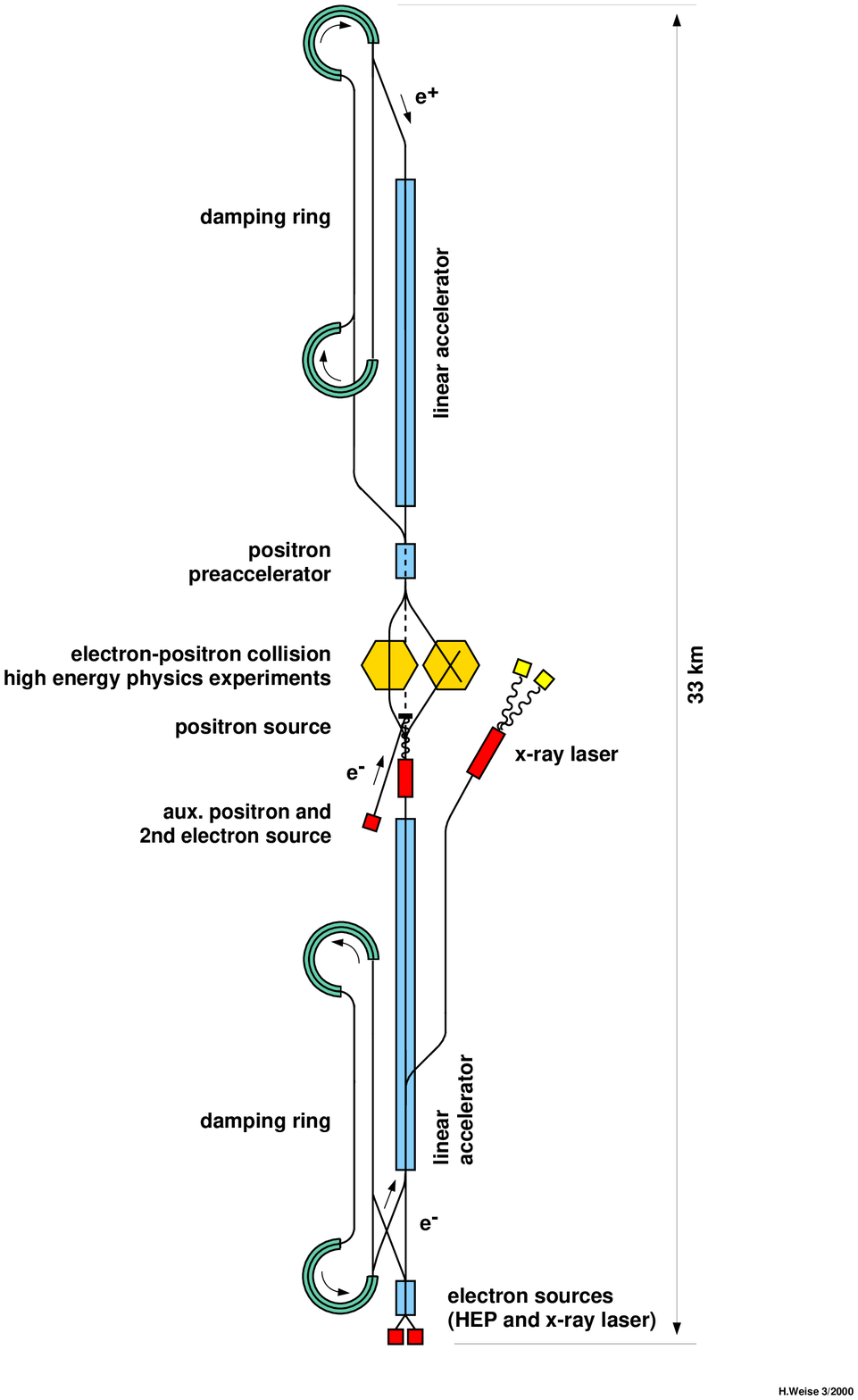,width=14cm,angle=0}
\caption{\label{ovw_layout}
Sketch of the overall layout of TESLA (the second interaction region with crossing angle is optional and not part of the baseline design.)}
\end{center}
\end{figure}

The overall layout of the TESLA linear collider and laser facility is shown in Fig.~\ref{ovw_layout}. The  total site is 33\,km long. The main components are a pair of
linear accelerators, one for electrons and one for positrons, pointing at each other. The first low energy part of the electron accelerator also provides the electron beam for the X-ray Free Electron Laser. Each linear accelerator is
constructed from about ten-thousand one-meter long superconducting
cavities.  Pulsed radio frequency (1.3\,GHz) electromagnetic fields are guided five times per second for the duration of one millisecond into the cavities to accelerate the particles. 

 Superconducting technology provides important advantages for a linear collider. As the power dissipation in the cavity walls is extremely small, the power transfer efficiency from the radio frequency (RF) source to the particles is very high, thus
keeping the electrical power consumption within acceptable limits
($\sim$100\,MW), even for a high average beam power.
The high beam power is the first essential requirement to obtain a high rate of electron-positron collisions. 

The second requirement is extremely small sizes of the electron and positron beams at the
interaction point (IP). The relatively low RF frequency of the TESLA linear accelerators 
is ideally suited for conserving the
ultra-small size of the beams during acceleration. When
a beam is accelerated in a linear accelerator, the charged particles induce
electromagnetic fields (so-called wakefields) which act back on the
beam itself and can spoil its quality by increasing the energy spread and the beam
size. As these wakefield effects decrease strongly with increasing distance between the beam and the surrounding cavity walls, wakefields are much weaker in the larger cavities of accelerators working at low RF frequencies than in smaller cavities operating at higher frequencies.

For the same reasons, the superconducting linear accelerator of TESLA is also extremely well suited to drive an X-ray Free Electron Laser (XFEL), which also requires 
an electron beam with large average power, high
bunch charge, small energy spread, and small beam size.

Due to the large aperture of the TESLA cavities the alignment accuracy of the cavity axis with respect to the beam is relatively easy to achieve.
Moreover the bunch separation of more than 300 ns permits orbit control by a fast feedback system.
 Such a feedback system will maintain
the beams in collision at the interaction point, making TESLA quite
insensitive to mechanical vibrations which could otherwise lead to a
serious reduction of the interaction rate.

The benefits of superconducting cavities have been known since the beginning of linear collider research and development. However, the accelerating fields achieved in the early 1990's were too low and the projected costs based on the then existing superconducting installations were too high for a collider facility. The main challenge for TESLA was therefore a reduction in the cost per unit accelerating voltage by a factor of 20.

Building on existing experience with superconducting cavities from CERN, CEBAF, Cornell, DESY, KEK, Saclay and Wuppertal, the TESLA collaboration succeeded in meeting the challenge:

\begin{itemize}
\item  By continued improvements of the base material (niobium), the cavity treatment, and the welding/assembly procedures, accelerating fields exceeding 25\,Million Volts per meter (MV/m) have been reliably achieved.

\item Through numerous design optimisations the costs per unit length of the
superconducting structures and the cryostat were reduced by a factor four for a large scale production.
\end{itemize}

\section{The Superconducting Cavities: Basic Elements of the Accelerator
}

\begin{figure}[h!tb]
\begin{center}
\epsfig{figure=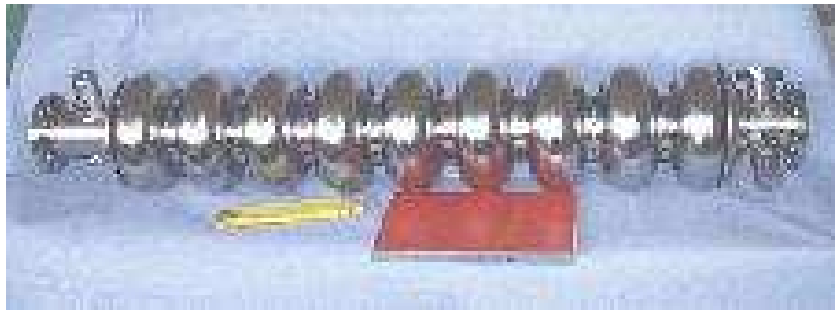,width=11.cm,angle=0}
\caption{\label{ovw_cavphoto}
The 9-cell niobium cavity for TESLA.}
\end{center}
\vspace{-0.cm}
\end{figure}

The basic elements of TESLA are 9-cell superconducting cavities, made from niobium (Fig.~\ref{ovw_cavphoto}) and cooled by
superfluid helium to $ -271$\,C. The operating frequency is 1.3\,GHz. For the chosen length of TESLA the accelerating field of the cavities (gradient) needed  
to reach a total collision energy of $ 500$\,GeV is $ 23.4$\,MV/m.

In order to demonstrate the technical feasibility of high gradients in superconducting cavities,
the TESLA collaboration began in 1992 to set up a test facility for superconducting cavities,
the TESLA Test Facility (TTF) at DESY. The infrastructure of TTF incorporated all the experience gained in the collaborating institutions, providing through all steps of cavity treatments a dust free environment, which  had been found essential in obtaining high gradients.

\begin{figure}
\begin{center}
\epsfig{figure=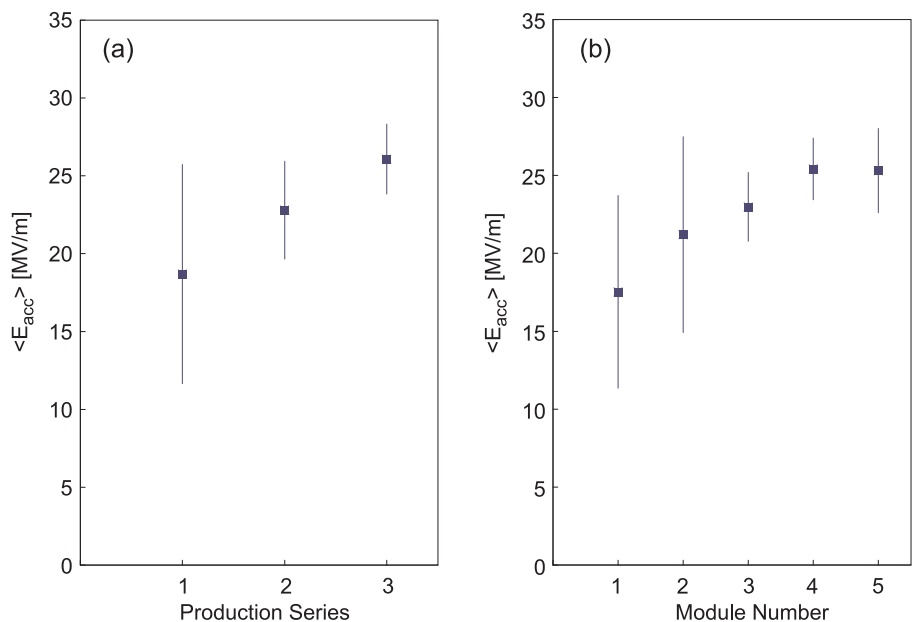,width=12cm,angle=0}
\caption{\label{ovw_cavperform}
Average accelerating gradients at a quality factor $Q_0>10^{10}$ measured in the
vertical test cryostat: \newline
(a) of the cavities in the three production series; (b) of 
the cavities installed in the first five cryogenic modules for TTF. The error bars give the standard deviation.
}
\end{center}
\end{figure}

To date, more
than sixty 9-cell cavities have been fabricated by industry and tested at TTF.
Fig.~\ref{ovw_cavperform} shows the average gradient obtained in the
years 1995 to 2000. The steady improvement of the average gradient over the
past years clearly indicates that the high performance cavities
required for the 500\,GeV collider can now be reliably produced.

Further progress in cavity performance has recently been obtained
by applying electro-polishing to the niobium surface. Test results with
single-cell resonators repeatedly show gradients well above 35\,MV/m:
The best single-cell performance obtained to
date is $ 42$\,MV/m. First results
for 9-cell electro-polished cavities show gradients above
30\,MV/m. 

\begin{figure}
\begin{center}
\epsfig{file=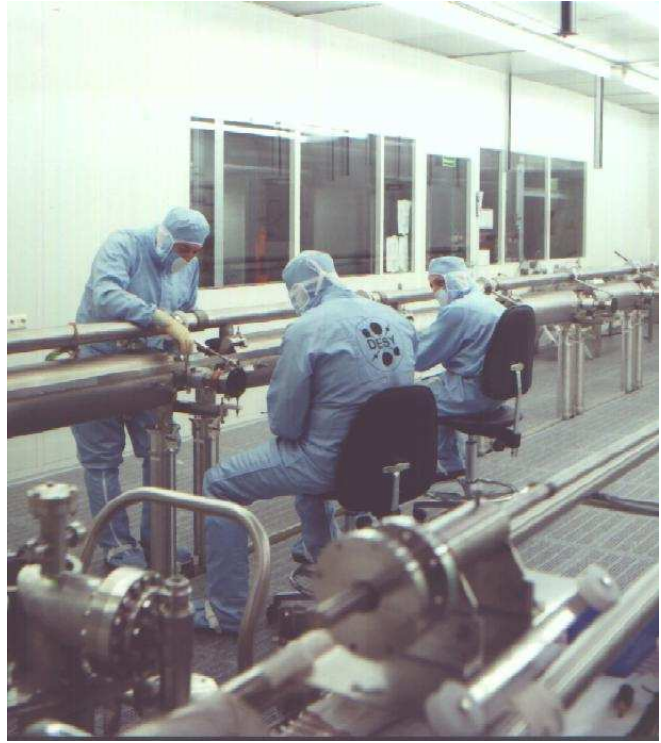,width=10cm,bb=-8 50 633 743}
\end{center}
\caption{\label{ovw_string}
Assembly of a string of eight 9-cell cavities in the clean room at TTF.}
\end{figure}

The TESLA collaboration also decided in 1992 to construct a short prototype linear accelerator (500\,MeV) as an integrated
system test, and to demonstrate that a linear collider based on superconducting cavities can be built and 
operated reliably. The 
commissioning and operation of the TTF linear accelerator has been the second essential milestone reached on our way to demonstrate the feasibility of the TESLA technology. The linear accelerator is constructed
from accelerator modules similar to those required for the collider. Each 12\,m long module
is comprised of a string of eight 9-cell cavities (Fig.~\ref{ovw_string}),
plus beam focusing and diagnostic
components.   We have so far tested three modules and operated the linear accelerator for more than 8600 hours.

\section{The Linear Collider Complex
}

The electron beam for the TESLA collider is generated in a polarised laser-driven source. After a short section of conventional normal conducting
linear accelerator, the beam is accelerated to 5\,GeV in superconducting structures
identical to the ones used for the main accelerator. The electrons are stored in a damping ring at 5\,GeV to reduce the beam size to values adequate for high luminosity operation. As the train of 2820 bunches is 300\,km long, a compression scheme is used to store the bunches in the damping ring. Reducing the distance between bunches by a realistic factor of 16 yields an 18\,km circumference. Using the so-called ``dog bone'' design with two 8\,km straight sections, most of the circumference can be placed inside the main accelerator tunnel. Only two 1\,km loops are needed at either end. After damping, the bunch train is decompressed and injected into the main linear accelerator.

A conventional positron source cannot provide the total charge of about
$5\cdot 10^{13} $ positrons 
per beam pulse, needed for the high luminosity operation of the TESLA collider. Therefore an intense photon beam is generated by passing the high energy electron beam through an undulator magnet placed after the main linear accelerator.  Positrons are produced
by directing the photons onto a thin target in which they are converted into pairs of electrons and positrons. After acceleration to 250\,MeV in a normal--conducting linear accelerator the positron beam is transported to a 5\,GeV superconducting accelerator, after which it is injected into the positron damping ring. This source can also generate a polarised positron beam.

The RF-power to excite the superconducting cavities is generated by some 300 electron tubes (klystrons) per linear accelerator. The required peak power per klystron is about
10\,MW. The high-voltage pulses for the klystrons are provided
by conventional power sources (modulators).

The cryogenic system for the TESLA accelerators is
comparable in size and complexity to the one currently under
construction for the Large Hadron Collider (LHC) at CERN. Seven cryogenic plants cool the linear accelerators to 2\,K. The cooling capacity of the first section of the electron accelerator is higher than the rest to accommodate the additional load from the  XFEL beam pulses.

The about 1.6 km long beam delivery systems between the linear accelerators and the collision point, where the experiment is located, consist of sections to remove the 
beam halo, beam diagnostics and correction, and the final focus system, consisting of magnetic lenses which focus the beams at the collision point down to spot sizes of about 550\,nm width and 5\,nm height.  The design of the final focus is essentially the same as the
Final Focus Test Beam (FFTB) system successfully tested at the Stanford Linear Accelerator Centre (SLAC) and the beam optics requirements of  TESLA  are comparable to those achieved at the FFTB. The beams can be kept in collision at the interaction point to a high precision by using a fast bunch-to-bunch feedback, which measures and corrects the beam-beam offset and crossing angle on a time scale comparable to the time between bunches. A prototype of the orbit feedback system has been installed
and successfully tested at the TTF linear accelerator. The beam delivery systems have been optimised for a single head--on
interaction point. The magnet systems and the
beam line layout are designed for a beam energy of up to
400\,GeV. 

\begin{figure}
\begin{center}
\epsfig{figure=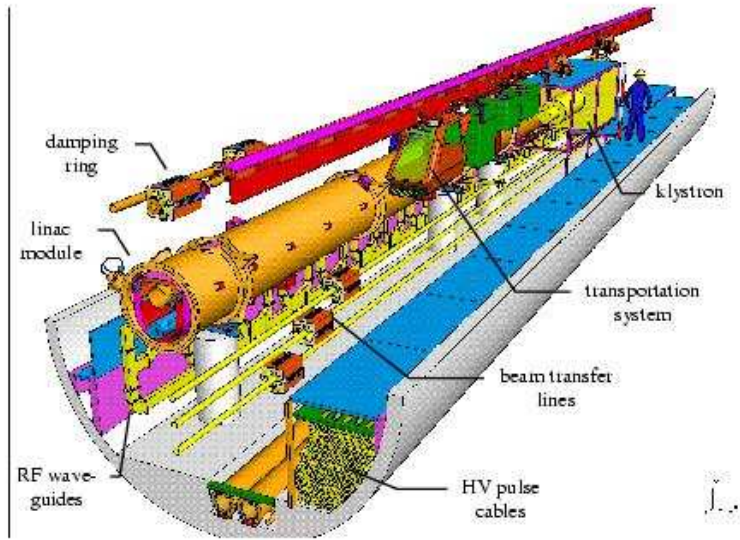,width=11cm,angle=0}
\caption{\label{ovw_tunnel}Sketch of the 5.2\,m diameter TESLA tunnel.}
\end{center}
\end{figure}

The two linear accelerators as well as the beam delivery systems will
be installed in an underground tunnel of 5.2\,m diameter (see
Fig.~\ref{ovw_tunnel}).
One experimental hall is foreseen to house the big high energy physics detector setup at the collision point; it  can be
extended for a second detector should a second interaction region be
constructed. The seven surface halls for the
cyrogenic plants are connected to the underground tunnel by access shafts.  The
halls also contain the power sources (modulators) which generate the high voltage pulses for the klystrons located in the accelerator tunnel, thus allowing access to the modulators during machine operation. The exchange of klystrons, however, will require an interruption of the machine operation. Assuming an average klystron lifetime of 40,000\,hours, a maintenance  day every few weeks will be necessary.

The baseline design for the 500\,GeV machine relies entirely on the components installed and tested at TTF, the
only exception being a further optimisation of the mechanical layout of the
modules. The modified module
improves the fill factor (ratio of active length to total length) and
hence increases the beam energy for a fixed accelerator length. A further improvement is expected from the so-called superstructure: in this concept multi-cell cavities are joined together to form  a single unit supplied by one RF power coupler, thus reducing the cost of the RF power distribution system. To be on the conservative side the parameters and cost estimate for the 500\,GeV
machine are based on the TTF-like accelerator modules, since tests
of the superstructure are still in preparation.

Some key parameters of the linear collider operating at 500\,GeV
are given in Table~\ref{ovw_parms500}, showing the very high luminosity which is provided by the TESLA concept based on low frequency superconducting cavities.

\begin{table} \begin{center} 
 \begin{tabular}{lcc}
 \hline
\addboth & & TESLA-500 \\
\hline
\hline
\addtop Accelerating gradient &  MV/m & 23.4   \\
 Total site length &  km & 33 \\
 No. of accelerating structures & & 21024 \\
 No. of klystrons & & 584 \\
 Klystron peak power & MW & 9.5 \\
 Repetition rate & Hz & 5  \\
 Beam pulse length &  $\mu$s & 950  \\
 No. of bunches per pulse &  & 2820  \\
 Bunch spacing &  ns & 337  \\
 Charge per bunch & $10^{10}$ & 2 \\
 Beam size at IP (width,height) &  nm & 553, 5  \\
 Bunch length at IP &  mm & 0.3  \\
 Luminosity &  $10^{34}{\rm cm}^{-2}{\rm s}^{-1}$ & 3.4 \\
 Power per beam &  MW & 11.3  \\
\addbottom Two-linac primary electric power &  MW & 97  \\    
\hline
\hline 
 \end{tabular}
\caption{\label{ovw_parms500}TESLA parameters for the $500$\,GeV baseline 
design. The machine
length includes a 2\% overhead for energy management. The klystron power and 
primary electric
power quoted include a 10\% regulation reserve. }
 \end{center}\end{table}

\section{Energy  Potential of the Linear Collider
}

 Higher beam energies than for the baseline design are possible within the site length since:

\begin{itemize}
\item The filling factor of the linear accelerators can be further increased by about 6\% by a modified design of the cavity structure (superstructures), and hence the maximum energy for a fixed accelerating gradient and site length.
\item The physical limit for the gradient in niobium structures at 2\,K
  is above 50\,MV/m, and several 9-cell cavities have already reached
  gradients around 30\,MV/m at TTF.  Electro-polishing, followed by
  low-temperature bake-out, has yielded systematically higher performance
  single-cell cavities, with gradients up to 42\,MV/m.
\item A method has been developed and successfully demonstrated at TTF which compensates the mechanical deformation of the cavities resulting from the strong electromagnetic fields. This active mechanical stabilisation using fast tuners, thus stabilizes the resonant frequency of the cavities and maintains an optimal power transfer from the RF generator.
\end{itemize}

Based on the described and ongoing progress in building high gradient
cavities we assume that TESLA will be built from the very beginning with
 superstructures and with cavities reaching gradients of on average 35\,MV/m, thus allowing a total collision energy of 800\,GeV. The
beam delivery system and the magnets in the linear accelerators have
been designed for a beam energy of up to 400\,GeV. \\ \\

 Collision energies above 500\,GeV can
be reached in two steps:
\begin{itemize}
\item
 A total energy of 650\,GeV can already be obtained
in the baseline design as the cooling plant capacity has a 50\% overhead, thus allowing the gradient in the cavities to be 
increased by 20-30\%.
\item
 In order to reach 800\,GeV at maximum luminosity
 two upgrades are required:  the cooling capacity of the cryogenic
plant must be increased  and the number of the RF stations must be doubled.
\end{itemize}

The positron production scheme maintains the required yield for high luminosity operation down to electron energies of 160 \,GeV, thus allowing high luminosity operation at the threshold of top quark pair production.

For high luminosity operation at the mass of the $Z$ boson, an
electron beam is extracted at 45 \,GeV after the first $\sim 3$\,km of
the linear accelerator and transported to the interaction region using
a bypass line. A second pulse is accelerated in the remainder of the
accelerator  and is used to produce the positrons.

\section{Second Interaction Region and Further Options
}

The baseline design of TESLA plans for a single particle physics interaction region.
Unlike a storage ring collider, a linear collider cannot serve several
interaction regions (IR) simultaneously with the same beam. It is possible,
however, to switch the beam between two experimental stations. The option of a second IR has been investigated. This IR can be used
for electron-positron collisions, with the same luminosity as the
primary IR (assuming that the so-called crab-crossing scheme is used). Unlike the primary IR, the
second IR will have a crossing angle of $\sim$34\,mrad, and is
therefore also suitable for the $e\gamma$ and $\gamma\gamma$ collider modes
of operation described in Part VI-1. 

Electron-electron collisions (at one or both of the IRs) can be
provided by reversing magnet polarities and adding an electron beam
source to the (nominal) positron branch of the collider. 
The expected performance for the $\gamma\gamma$ and $e^-e^-$ modes of operation are included in the discussion of machine parameters.

In addition to the collider operation, TESLA also offers options for fixed
target physics. It is possible to accelerate (in parallel with the
main collider beam) a low-intensity spin-polarised electron beam which
can be deflected into a separate beam line and used for a polarised
target experiment (see Part VI-3).  Except for the additional
experimental beam line and low-current polarised electron source, the
impact on the accelerator itself is marginal, since the required
additional RF-power is far less than 1\% of the nominal power.

The first part of the electron linear accelerator
can be used as an injector for the electron ring of the existing HERA collider, which could be operated as a pulse stretcher to deliver a continuous beam at
15 to 25\,GeV for fixed target nuclear physics experiments.  This option provides a beam with properties very similar to a recent design worked out at CERN (see Part VI-4).
 
As another option, the TESLA tunnel can be connected to the straight section West of the HERA proton ring, allowing the
electron beam from TESLA to collide with protons stored in HERA (see Part VI-2).  

 These additional possibilities for experiments at TESLA are not part of the baseline design and have therefore not been evaluated at the same level of detail as the collider and X-FEL.

\section{
The X-ray Free Electron Laser (XFEL)
}
  
The conservation of the high beam quality during
acceleration also makes the TESLA linear accelerator an excellent
driver for an X-ray Free Electron Laser (XFEL).
The X-ray Free Electron Laser functions by passing an electron 
beam pulse of small cross section and high peak current through 
a long periodic magnet structure (undulator). The interaction of 
the emitted sychroton radiation with the electron beam pulse 
within the undulator leads to the buildup of a longitudinal 
charge density modulation (micro bunching), if 
a resonance condition -- determined by the electron beam energy and 
the undulator period -- is met. The electrons in the 
developing micro bunches eventually radiate coherently and 
the number of emitted photons grows exponentially. 
This is the basic principle of a Single-Pass Free Electron Laser
(FEL) operating in the Self-Amplified-Spontaneous-Emission (SASE)
mode. The concept of using a high energy electron linear accelerator
for building an X-ray Free Electron Laser (FEL) was first proposed for
the Stanford Linear Accelerator.

\begin{figure}[tbh]
  \begin{center}
\epsfig{file=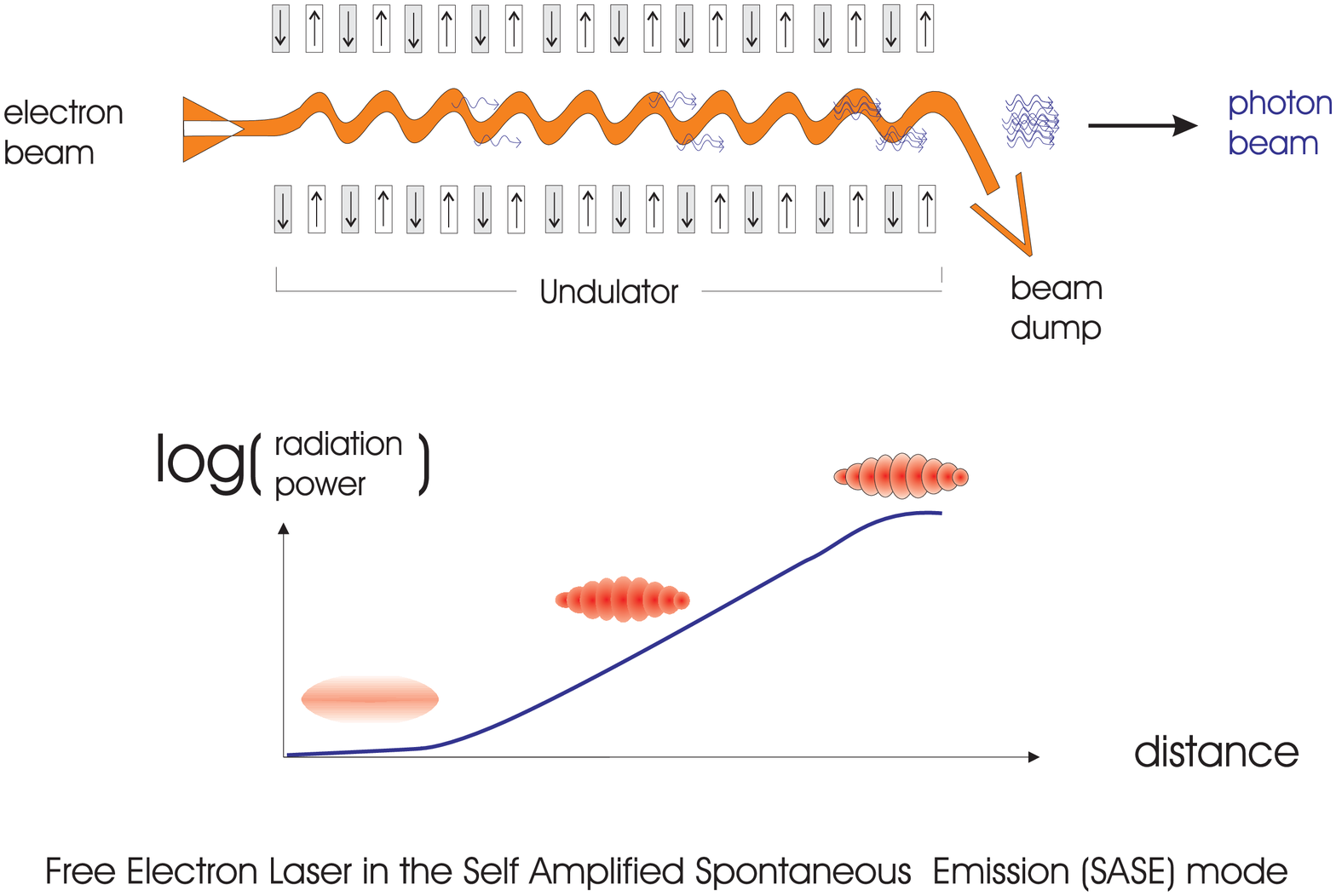,height=8cm}
  \end{center}
  \caption{\label{fel-layout}
Schematic Diagram of a Single-Pass Free Electron Laser (FEL) operating in
the Self-Amplified-Spontaneous-Emission (SASE) mode. The bunch density 
modulation (micro-bunching), developing in parallel to the radiation power, 
is shown in the lower part of the figure. Note that in reality the number of slices is much larger.}
\end{figure}

The principle is shown schematically in Fig.~\ref{fel-layout}. The bunch density modulation (micro-bunching), developing in parallel to the radiation power, is schematically shown in the lower part of the figure. 
The SASE-FEL does not require the optical cavity resonator normally used in multi-pass Free Electron Lasers, working at longer wavelength. It can therefore deliver radiation with a wavelength down to X-ray wave lengths, where mirrors no longer work.

\begin{figure}[tbh]
  \begin{center}
\epsfig{file=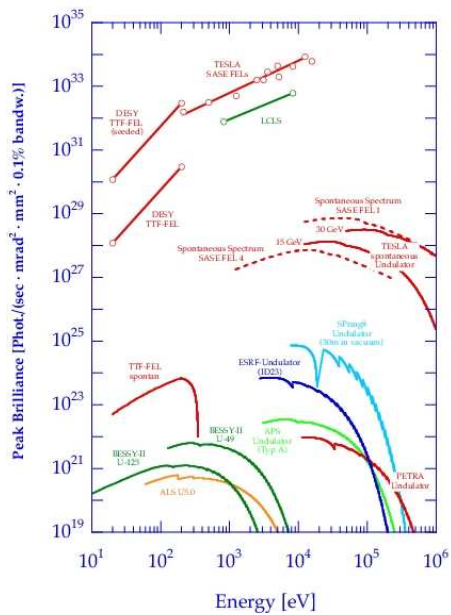,height=10cm}
  \end{center}
  \caption{\label{fel-peak}
Spectral peak brilliance of XFELs and undulators for spontaneous radiation
at TESLA, together with that of third generation synchrotron radiation
  sources. For comparison, the spontaneous spectrum of an XFEL
  undulator is also shown. 
}
\end{figure}

The SASE FEL delivers coherent X-ray pulses of
about 100 fs length, compared to about 30 ps at state-of-the-art synchrotron radiation sources. As shown in Fig.~\ref{fel-peak} the peak brilliance per pulse exceeds present sources by eight or more orders of magnitude.

As the SASE FEL does not depend on atomic excitation levels,
it can be tuned over a wide range of wavelengths. This tunability was
recently demonstrated at the TESLA Test Facility in the wavelength
range between 80\,nm and 180\,nm.

To achieve photon wavelengths as small as atomic dimensions ({\AA}ngstr{\o}m), an electron beam of about \,30 GeV is required.
Since the X-ray FEL concept represents a considerable extrapolation of
present day FEL technology, a
test of the SASE FEL concept at the TTF, in a wavelength region previously inaccessible ($\lambda\approx 100$\,nm) has been performed. Lasing was first observed in February 2000, and a number of experimental studies at the TTF-FEL have since been carried out. An upgrade of the TTF linear accelerator to 1\,GeV beam energy is in preparation, allowing the FEL to reach 6\,nm wavelength. The  facility will provide  operational and scientific experience for the proposed X-ray FEL laboratory.

\begin{figure}[tbh]
  \begin{center}
\epsfig{file=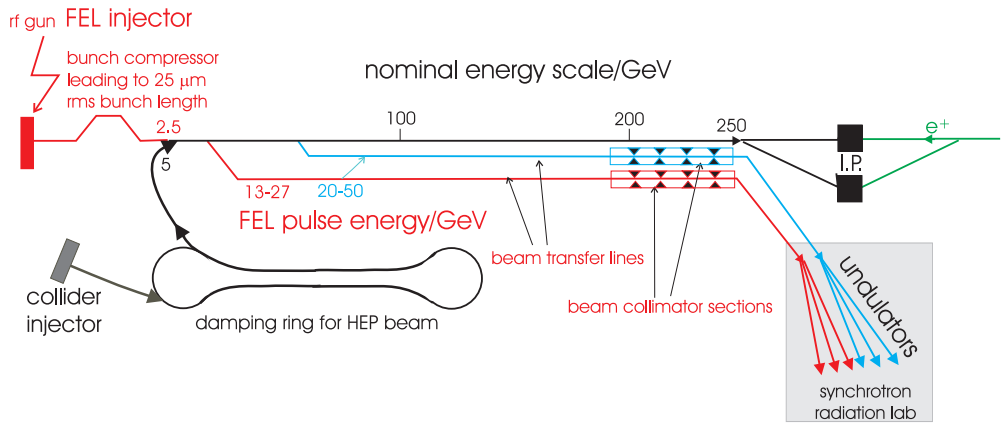,height=5.5cm}
  \end{center}
  \caption{\label{fel_laboratory}
Sketch of a coherent X-ray source laboratory based on the TESLA linear
collider installation. Two electron beam lines, extracted from the electron linear accelerator at energies between 11 and 35\,GeV,
are guided inside the accelerator tunnel to the X-ray laboratory.}
\end{figure}

Fig.~\ref{fel_laboratory} shows a sketch of the laboratory as part of the overall layout of TESLA. The XFEL electron beam is
provided by a special injector. The first 3 kilometers of the electron accelerator are used to accelerate the beam which drives the X-ray FEL laboratory. Two extraction
beam lines are planned for supplying electron energies typically between 13 and 35\,GeV. About 900 meter in front of the XFEL laboratory, the beams are deflected into a 4 m diameter tunnel, separating the beams from the collider tunnel. The extraction lines can be used in parallel, so that various beam energies are available in the X-ray laboratory. 

\begin{figure}[hbp]
  \begin{center}
\epsfig{file=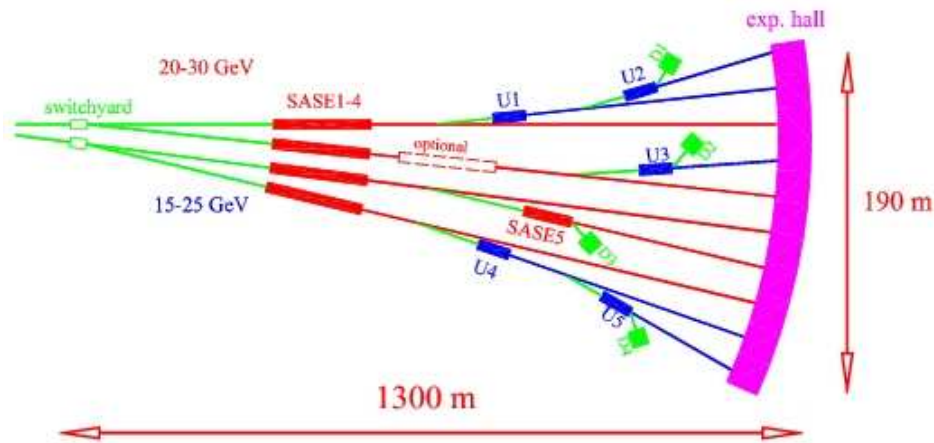,height=6cm}
  \end{center}  
  \caption{\label{fel_switchyard}
Beam switchyard distributing two electron beam lines of different energies to various undulators. SASE1 through SASE5 are FEL undulators while U1 through U5 are undulators for spontaneous radiation. 
}
\end{figure}

At the end of the extraction tunnel, a beam switchyard distributes 
the beams to different undulators, as shown in Fig.~\ref{fel_switchyard}.
The quality of the electron beam behind the FEL is still very good, so that
it can be passed through another FEL or through further undulators generating spontaneous radiation.

The electron pulse structure consists
of trains of electron bunches with a repetition rate of 5 Hz. The accelerator will run in a mode where pulses for high energy collisions
and the XFEL alternate. This means that the first part of the electron
linac, which serves both the XFEL and the collider, operates at a rate
of 10 Hz. Both user communities can define the beam properties like
energy, emittance, current, time structure, etc., to a very large
extent independently of each other, since separate injectors will be
used, and the RF system allows for changes to power and pulse length
from pulse to pulse.

The FEL laboratory building is located at the new campus close to the
collider interaction region. The layout of the laboratory follows a
similar approach as used for the third generation light sources: For
the baseline design, the plan is to build five laser beam lines, each
equipped with three experiments, and five beam lines for spontaneous
radiation, each with one experiment.

\newpage
In Table~\ref{tab:fel_parameters} some key parameters of the XFEL are summarised.
\begin{table}
\begin{center}
  \begin{tabular}{p{10cm}lc}
\hline
\multicolumn{3}{l}{\addtop Linac Parameters}\\
\hline\hline
Optimised gradient for XFEL operation &MV/m &18\\
Linac repetition rate for XFEL &Hz &5\\
Bunch length (rms) &fs& 80\\
Bunch spacing& ns& 93\\
Number of bunches per train&& 11500\\
Bunch train length & $\mu$s& 1070\\
Bunch charge &nC& 1\\
Over-all power efficiency AC to electron beam&  \% &30\\
\hline
\multicolumn{3}{l}{FEL Parameters}\\
\hline
Energy                       &  keV    & 0.5 - 14.4\\
Wavelength                   &  $\AA$  & 25 -0.85\\
Saturation length            &  m      & 60 - 220 \\
Peak power                   & GW      & 100 - 37 \\
Average power                & W       & 550 - 210 \\
Photon divergence (FWHM)    & $\mu$rad& 6 - 0.8 \\
Beam diameter/ experiment (FWHM) & mm & 1 \\
\addbottom photons/pulse                & $10^{12}$ & 10 - 2\\
\hline
\hline
\end{tabular}
\end{center}
\caption{\label{tab:fel_parameters}
{ Parameters for the XFEL operation.}}
\end{table}

\chapter{Project Cost Estimate and Schedule } 
The costs for the TESLA project are given separately for the 500\,GeV
electron--positron linear collider, the X-ray Free Electron Laser
(incremental cost for accelerator, beam delivery and the X--ray
laboratory), and one detector for particle physics. All costs are
given in Euro and in year 2000 prices.

\section{The Linear Collider}

\begin{figure}[b!h]
  \begin{center}
    \nfigure{\includegraphics[height=7cm]{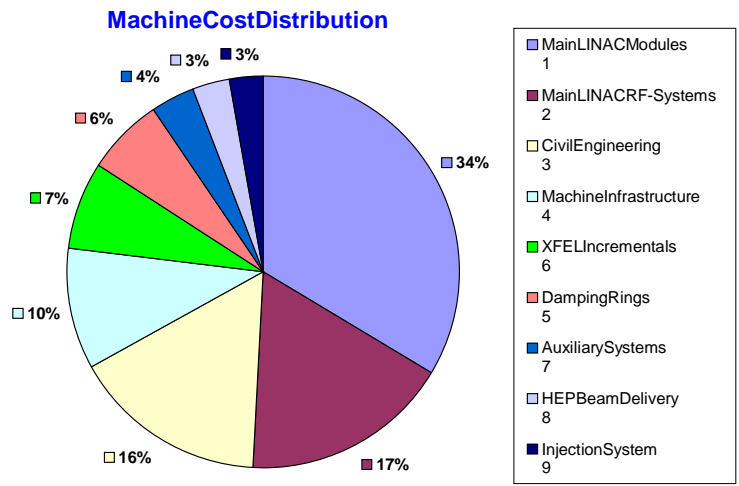}}
  \end{center}
  \caption{\label{tesla_cost2} Distribution of the accelerator
    sub--systems in percent of the total cost. The costs of the X--ray 
FEL laboratory and the detector for particle physics are not included here.}
\end{figure}

\begin{figure}[p!]
  \begin{center}
  \nfigure{\includegraphics[height=14cm,angle=+90,bb=0 244 575 598]{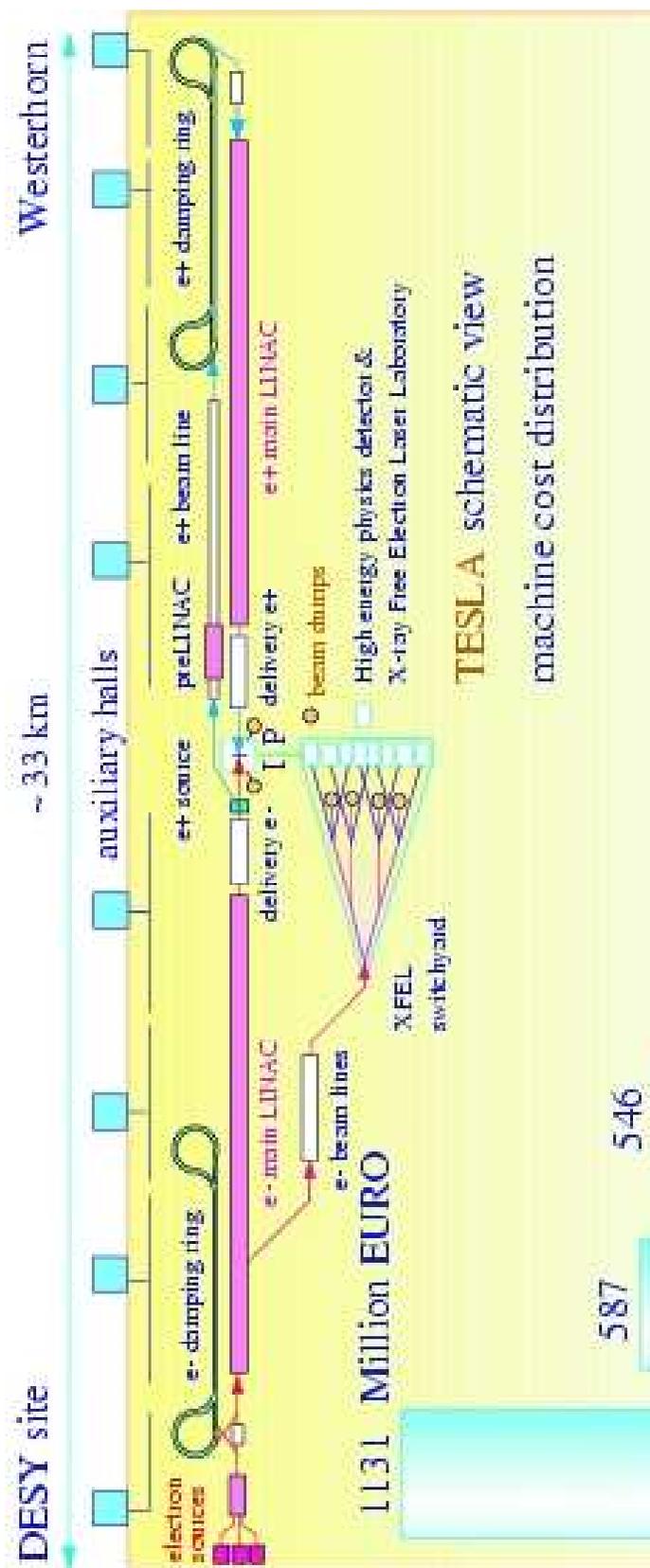}}
  \end{center}
  \caption{\label{tesla_cost1}Overview of the accelerator investment costs.}
\end{figure}

The cost for the 500\GeV\  linear collider baseline design with one
interaction region is
\begin{itemize}
\item   \totalcost\  million \Euro.
\end{itemize}
The cost for the major subsystems is detailed in 
 Fig.~\ref{tesla_cost2}, Fig.~\ref{tesla_cost1} and table~\ref{cost}. 

\clearpage
\begin{table}[p!]
\begin{center}
  \begin{tabular}{lp{7cm}r} 
\hline
Sub--system & Component included  & \multicolumn{1}{c}{cost} \\
 & & \multicolumn{1}{c}{million \Euro} \\
\hline\hline
Main acceleration structure &   {cavity string, input
couplers, modules, HOM couplers, tuning system, cryostats,
s.c.~quadrupoles, steering magnets, instrumentation} & \linea\\
\hline
Main linac RF system &  
{RF power supplies, modulators, HV pulse cables, transformers,
klystrons, wave guide system, low level RF controls, interlocks and
cables} & \lineb\\
\hline
Injection system &    
{RF gun system, accelerator modules and RF system for 5\,GeV linac for
electrons and positrons, positron source, conventional
pre--accelerator, beam transfer lines, diagnostics} & \linec \\
\hline
Damping ring &    
{magnet--, vacuum--, RF--system, power supplies, beam instrumentation,
bunch compressor, injection and ejection systems} & \lined \\
\hline
Beam delivery system of LC &  
{beam transport lines, beam collimation, beam dump systems, vacuum
systems, power supplies, feedback systems, diagnostics} & \linee \\
\hline
Civil engineering &   
{33\,km tunnel, surface buildings, connecting shafts, experimental
hall for high energy experiment, damping ring tunnels, beam dump  
halls} & \linef \\
\hline
Infrastructure &   
{tunnel infrastructure, cable trays, power distribution, main power
connection, cryoplants and cryogenic distribution system, ventilation,
test facility, water cooling system} & \lineg \\
\hline
auxiliary systems &  
{control systems, vacuum pump stations, cabling, interlocks, magnet 
supplies, miscellaneous} & \lineh \\
\hline\hline
\end{tabular}
\caption{\label{cost} 
The cost for the subsystems of the TESLA 500 GeV collider.}
\end{center}
\end{table}
\clearpage
The cost goal of  2000\$/MV   
 for the complete accelerating modules (including superconducting
 cavities, power couplers, cryostat, quadrupoles etc.), set by
 Bj{\o}rn H. Wiik for the TESLA collaboration in 1992, has been met to
 within \gradientpercentage\%.

\vspace*{-4mm}
\subsection*{Capital Cost Estimate Basis}
\vspace*{-4mm}

The cost estimates for all major components have been obtained from
studies made by industry, and are based on a single manufacturer
supplying the total number of a given component. A schedule of three
years peak production  plus one year for start--up for the total number
of each component was specified. The four--year production cycles
required for the various components are scheduled within the total
construction time of eight years. The schedule  was considered
feasible by the companies involved in the study.

A production period of four years corresponds to an average
production rate of 32\,m of machine per working day. The
corresponding numbers for the proton storage ring of the Hadron
Electron Ring Accelerator HERA at DESY were about 25\,m/day; for the
Large Hadron Collider LHC at CERN about 40\,m/day are planned.

Several collaborating institutes were in charge of evaluating parts
of the cost.
A planning group consisting of the persons responsible for each of the
major subsystems, together with experienced senior scientists from the
collaboration has been continuously reviewing the technical layout of
the system and the cost evaluations. The procedures which were
followed in the cost evaluation for the major components can be found
in Part~II. 

No additional contingency has been added, since the last two major
particle physics facilities in Europe -- the Large Electron Positron
collider LEP at CERN and HERA -- were built within budget and on
time.  
\vspace*{-4mm}
\subsection*{Manpower Requirements}
\vspace*{-4mm}
The personnel needed for the different stages of the project --
design, procurement, fabrication and assembly, testing, installation
and commissioning -- has been estimated mainly on the basis of the
experiences gained at the TESLA Test Facility and in large projects
like HERA.
A total of \personyears~person--years will be required. It is assumed that
this manpower will be supplied by the collaborating institutes. The
manpower needed for manufacturing of components in industry  is
included in the investment costs.

\vspace*{-4mm}
\subsection*{Maintenance and Operating Cost}
\vspace*{-4mm}

The operating costs include the  electrical power, the regular
refurbishing of klystrons, water, and the helium losses. The numbers
are determined assuming current prices and an annual operation time of
5000 hours.
Costs for general maintenance and repair have been estimated assuming
2\% per year of the original total investment costs, corresponding to
the DESY experience. In total, the costs  for the operation of the
accelerators has been estimated at \operating~million \Euro\  per year.
For critical components (such as accelerator modules) a number of
spares will be produced; these costs have been included in the
investment costs.
\vspace*{-4mm}
\subsection*{Time Schedule}
\begin{figure}[p]
  \begin{center}
   \hspace*{-2mm}\includegraphics[height=19cm,bb= 1 114 553 786]{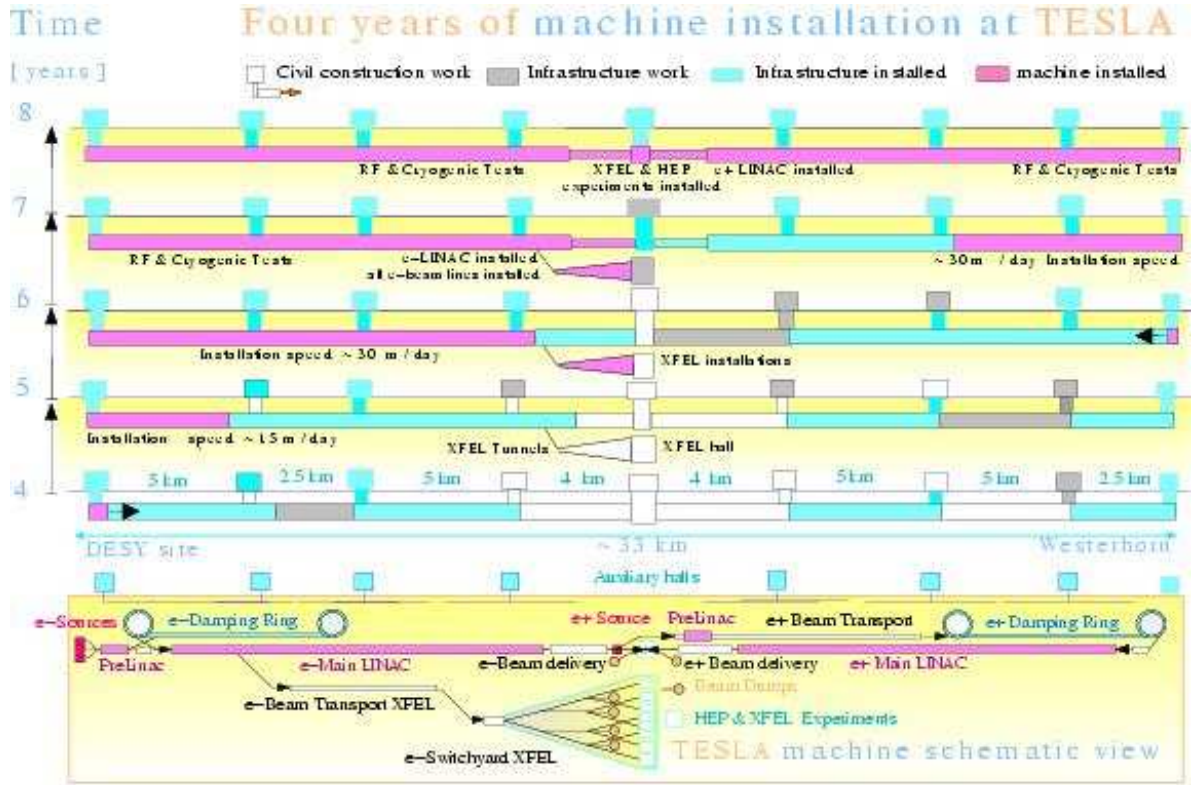}
  \end{center}
  \caption{\label{tesla_construction}Time schedule of work for 
  civil construction (bottom) and for the installation work 
  (top).
  Tunneling starts from four points along the tunnel in
  parallel and proceeds towards the interaction region. Civil
  construction, infrastructure work, and accelerator installation
  happen concurrently at different regions in the tunnel.
}
\end{figure}

\vspace*{-4mm}
The  construction time of TESLA  is 8 years. The estimate is based on
industrial studies and the experience gained from the construction of
HERA and the TESLA Test Facility. In Fig.~\ref{tesla_construction} a and
b the construction and installation schedule is shown, indicating the 
major activities. For details see part~II.

\section {Additional Cost of the XFEL Laboratory} 

The additional costs of the accelerator components required for 
the X--ray FEL amount to
\begin{itemize}
\item  \xfelacccost\   million \Euro,
\end{itemize}
 the dominant part being the civil engineering of the FEL experimental
 hall and the beam distribution system.
The equipment cost for the undulators, beam lines and experiments,
 including infrastructure -- 5 laser beam lines, each equipped with 3
 experiments, 5 other beam lines with 1 experiment each -- is
 estimated at
\begin{itemize}
\item \xfellabcost\  million \Euro.
\end{itemize}
The incremental cost of the XFEL--laboratory is listed in table~\ref{xfelcost}.
\begin{table}[h!] 
\begin{center}
  \begin{tabular}{lp{7cm}r}
\hline
Component  & Comment & million \Euro\\
\hline\hline
incremental cost for XFEL & 
{RF gun, 500\,MeV linac, upgrade of 50\,GeV linac to 10\,Hz (RF system
and cryo system), bunch compressors, FEL and LC beam merging and 
separation, collimation, beam transport,electron beam abort system,
civil construction, diagnostics} & \xfelacccost\\
\hline
XFEL laboratory &  
{Undulators, beam lines, experiments} & \xfellabcost\\
\hline\hline
\end{tabular}
\end{center}
\caption{\label{xfelcost}Incremental cost for XFEL} 
\end{table}

\section{A Detector for Particle Physics}

The physics requirements have led to the design of a detector with a
high magnetic field and a calorimeter of unprecedented granularity,
as described in part~IV. 
The cost estimates of detector parts are based on the experience with
detectors already built  or presently under construction for the LHC
experiments.

Depending on the choice of technologies, the detector
is priced between
\begin{itemize}
\item \detcostone\  and \detcosttwo\  million \Euro,
\end{itemize}
 including data acquisition
system and detectors to measure beam energy and polarisation.
The price bracket is mainly due to the different approaches to
the electromagnetic calorimeter. The final price can only be
decided after intense further development work, more simulation
studies and after the collaboration has formed.

\chapter{TESLA as an International Laboratory}

\section{Basic Considerations}

Endeavours of the size and complexity of TESLA should be realised as truly international projects.  

The OECD Megascience Forum report  on "Particle Physics" of 1995 distinguishes four organisational models for future accelerators:
\begin{enumerate} 
\item National and regional facilities are built, financed and
operated by the host country or host region. Planning,
project--definition and definition of parameters are done in
international co--operation.

\item 'HERA--model': Large facilities which are not financed by one
country or one region alone. The host country or host region receives
contributions -- mostly in kind -- from participating countries or
institutions. Planning and project--definition are done in
international co-operation. The host country or institution is
responsible for the operation.

\item Very large projects where construction and operation are
realised through contributions in more equal shares by the
participating countries or institutions. The partners contribute
through components or subsystems in a similar way as large
collaborations in particle physics 
are building jointly a major detector facility. A
facility under this model would be the common property of the
participating countries or laboratories. These would also share the
responsibility and cost for operation.

\item Very large projects in the frame of an international organisation like CERN.
\end{enumerate}
Model 2 was very successfully implemented for the construction of HERA
at DESY, where the overall costs (excluding the DESY-infrastructure)
were financed to 78\% by the German Federal Government and the State
of Hamburg, while major hardware contributions were made by
collaborating countries.

For the TESLA project a larger international participation than for
the HERA--project will be required. 
A possible approach is that of a 'Global Accelerator Network', 
which is based on model 3 and has the following features:
\begin{itemize}
\item The project is open for participation of international 
and national research and academic institutions. 
\item The project would be part of the national programs of the
participating countries.
\item The capital investments would be made under the responsibility
of the participating institutions.
\item The participating laboratories and institutes would be able to
maintain and foster the scientific and technical in--house
culture. They would remain attractive for young scientists and
contribute at the same time to and participate in large, unique
projects.
\item The accelerator facility would be maintained and run to a large
extent remotely from the participating laboratories, using the modern
tools for communication and controls.
\item This approach would effectively combine world--wide competence,
ideas, manpower and financial resources.
\item The site selection would become a less critical issue.
\end{itemize}
In order to make optimal use of experience, manpower and
infrastructure, the accelerator built according to this model should
be put close to an existing laboratory. The site selection would lead
to specific financial obligations for the host country.

The 'Global Accelerator Network' is presently being investigated by
the International Committee for Future Accelerators (ICFA), which has
set up two specific working groups, one dealing with the general
aspects of the model, the other with the specific questions of remote
operation of accelerators. The recommendations and conclusions of
these working groups are to be published in 2001.

The model described above is in principle applicable for the entire
project and can assure for all parties involved the proper representation
in the decision making processes. Special attention must be paid to
balance the needs of the particle physicists and the users of the
lasers.

The international nature of the project has to be reflected in its
organisational structure. The international partners have to be fully
and formally involved in all decision--making processes, as it is their
project and not a project of the host country to which they
contribute. 

The concept of a TESLA project organisation with its own legal identity
is considered to be the most appropriate and adaptable to the needs of
the project. DESY offers to be the host laboratory for TESLA.

\section{TESLA as Project of Limited Duration}

Taking into consideration the scientific program as outlined in this
report, the initial project duration should be 25 years, including 8
years of construction. After 10 years of operation a possible
extension of the project should be decided upon.

\section{Project Contributions}

Each partner assumes responsibility for a specific component. The
components are designed, built, tested, installed, operated and
maintained under the responsibility of the respective partner. This
responsibility for specific components is maintained throughout the
project duration and includes also further developments. This allows
national money to be spent according to the national rules and not as
contribution to an international organisation which spends it
according to its rules. It also makes sure that the volume of the
financial involvement is directly linked to the technical
responsibility of each  partner. In addition contributions 
in manpower are possible and welcome.

It might be necessary that a small fraction of the project cost would
be paid from a Common Fund to which each partner makes a financial
contribution. 

The cost of operation of the facilities will be shared among all
partners according to a pre--defined procedure.

The mode of financing of the 
 participating laboratories and institutes, including the host,
would remain unchanged and their
staff would remain employees of their individual institution.

\section{International Project Convention}

TESLA as an international project can only be realised on the basis of
a long term involvement of the participating institutions and 
countries, which secure
the financing and operation of the project. Basis for the project
should therefore be an agreement between the participating countries
or institutions (Project Convention) in which they declare their intention
\begin{itemize}
\item to construct and operate TESLA,
\item to finance the project,
\item to set up a project organisation, and
\item to define the basic rules (e.g. financing, purchasing, and settlements of disputes).
\end{itemize}
In view of the long--term and substantial involvement this convention
should be signed by the governments of the participating countries or 
by bodies to whom the governments delegate their authority.
Alternatively it could be signed by the participating institutes.

\section{Legal Structure of the Project}

\begin{figure}[tb]
  \begin{center}
    \includegraphics*[height=10cm]{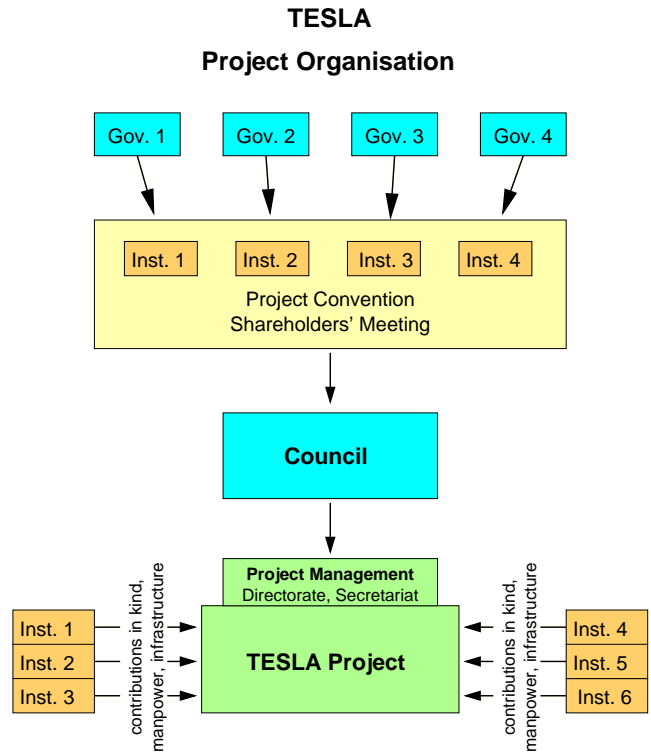}
  \end{center}
\caption{\label{executive_organisation}TESLA project organisation}
\end{figure}

As shown in Fig.~\ref{executive_organisation} all partners form a
shareholders meeting. Following the example of the ESRF the
governments are either represented themselves in this meeting or they
delegate their membership to one of their national participating
institutes. 
The shareholders' meeting elects a council, which is the supervisory
body for the project management.  This basic structure is common
practice in international scientific co-operation.

The project management will be responsible for the strategic planning,
project co-ordination, external relations and supervision of tasks
carried out by the participating partners or industries. It will be
supported by a co-ordination team, a secretariat and project groups,
responsible for their respective sub--projects.

The influence of each participant in the decision--making has to
reflect the importance of its contribution. However, smaller partners
will only identify themselves with the project if an effective
minority--protection is organised. Here the rules of the ESRF can serve
as an example. If the importance of the contribution is measured in
shares,  simple majority at the ESRF means at least 50\% of the shares
and the opposition of not more than half of the votes, each partner
having one vote; unanimity means 2/3 of the shares and no opposition --
abstentions always being possible.

The structure described above can be implemented in any national legal
system, in Germany it would be a Gesellschaft mit
beschr\"ankter Haftung, GmbH, which corresponds to a Limited Liability
Company in English and American Law.

\section{The Specific Role of the Host Laboratory}

In order to make optimal use of experience, manpower and infrastructure, the accelerator built according to this model should be put close to an existing laboratory. In this case, the host lab will make its infrastructure, personnel and services available to the project. 

A well defined legal relation has to be established between the
project management and the host laboratory. This can be done on the
basis of a management- and service-contract, whereby the host
laboratory assumes the responsibility for e. g.
\begin{itemize}
  \item security, radiation safety
  \item site-- and facility management
  \item technical infrastructure
  \item guest--services
  \item coordination of external communication/ media service.
\end{itemize}
The contract between the project management and the host laboratory must clearly define the respective responsibilities and obligations.

To ensure a close and smooth co-operation between the project management and  the host laboratory a steering committee is proposed, with the directors of the project and of the host laboratory as members, chaired by the project-chairman and co-chaired by the chairman of the host laboratory.

\chapter{Organisational Methods and Tools}

\begin{figure}[tb]
\begin{center}
  \includegraphics[height=12cm]{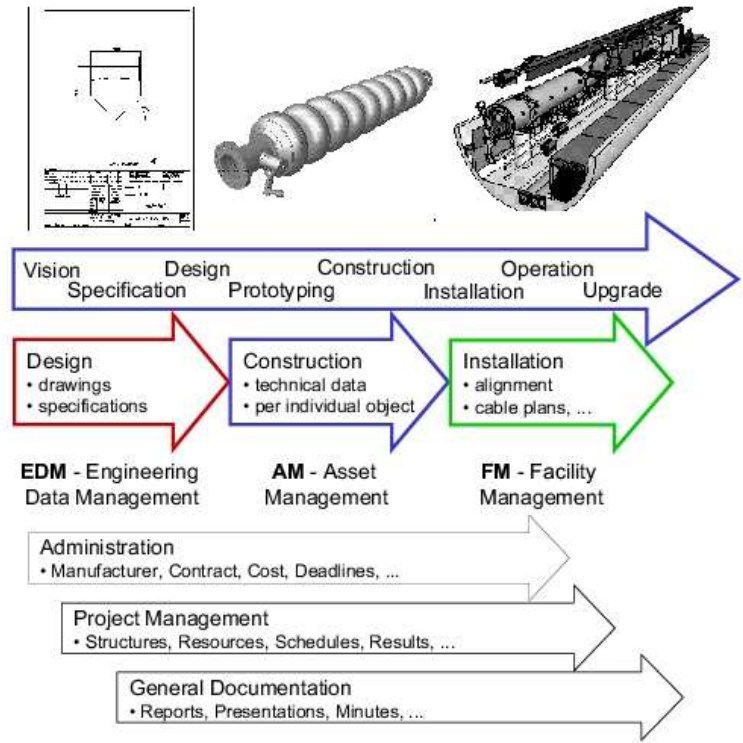}
  \caption{Tasks supported by information management during the TESLA                                           project. \label{fig:orgtool}}
\end{center}
\end{figure}

From an engineering point of view, TESLA shares many features with
other large--scale projects. With major contributions from many
countries or institutions, TESLA needs a well defined project
organisation and supporting methods and tools. Resources and
responsibilities, also with respect to documentation, change
management, and quality control, will be defined by the TESLA
collaboration. Based on the experience with LHC at CERN and with
large-scale projects in industry, the TESLA linear collider has to be
described in its systems and subsystems. The full life cycle of TESLA
has to be supported with consistent data. This includes design,
construction and installation, commissioning, and operation of TESLA
in the proposed 'Global Accelerator Network'. The present section
summarises the methods and tools for information management, which are
regarded as the technical basis for the project management at TESLA
(see Fig.~\ref{fig:orgtool}).

During the design, 3D mechanical CAD models of parts and complete
systems, as well as 2D projections of these items, very likely from
more than one mechanical CAD system, will be produced. These documents
are used in all later phases of the project. Electronic CAD
schematics, layouts, part lists, and many less strictly structured
documents like preliminary construction and installation procedures,
project plans, schedules and structures, contracts, meeting minutes
etc. will be produced as well and need distribution. In terms of
information management tools, the documents mentioned are accommodated
in an Engineering Data Management (EDM) system. Such a system is
evaluated in a pilot project at the TESLA Test Facility.

During construction and installation, information records per
individual object, referring back to the appropriate design document
will be collected. Quality control and documentation are main
issues. The construction phase may provide feedback to the design
documents yielding new versions and even unique versions for some of
the individual objects. The documents of this phase are mainly
accommodated in an Asset Management (AM) system for the individual
objects and in a Facility Management (FM) system for the larger
assemblies and for the buildings with their technical infrastructure
as a whole. For both systems the requirements are set up at DESY and a
pilot AM system is evaluated for information technology related
assets.

In the operation phase, access to all of the documents mentioned above is needed. In addition, this phase requires efficient support for troubleshooting and maintenance. Documents include a complete object history in the information records per individual object, describing operational performance, faults, and replacements, detailed operation and maintenance procedures with associated workflow support, as well as relations of the individual objects' construction parameters to the accelerator control system parameters, maybe in the form of automated electronic interfaces. Possible future upgrade programs of TESLA will require access to the design documents for doing major revisions.

Apart from the EDM, AM, and FM systems, the organisation also requires standard project planning tools as well as a close connection to the administrative (financial) information management systems. In addition, the design data from the various CAD tools used must be viewed and checked for consistency by many persons in a CAD tool independent way, through inexpensive Web interfaces. The installation and maintenance personnel need supporting devices like barcode scanners for parts identification integrated with mobile computers for remote access to the information management systems as an example. The automated interface between EDM and the administrative standard software (SAP) in use at DESY, the viewing tool, and a barcode system are evaluated in pilot projects.

\chapter{Site Considerations for the Construction of TESLA}

The TESLA facility should be built at an existing High Energy Physics laboratory in order to reduce project costs and construction time. In the Conceptual Design Report of 1997 both DESY and Fermilab have been considered as possible sites.

\begin{figure}[p]
  \begin{center}
    \includegraphics[height=18cm]{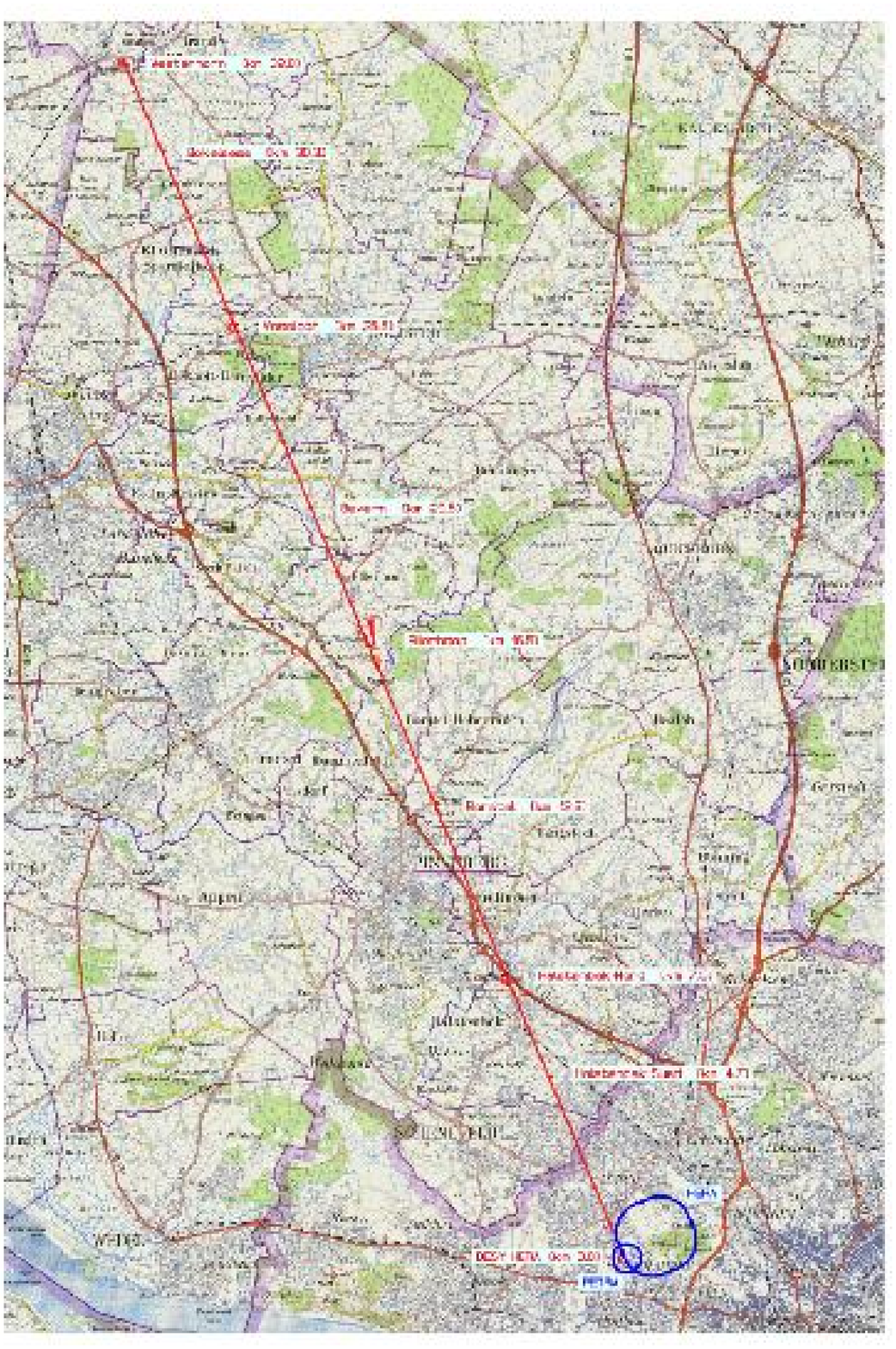}
  \end{center}
\caption{\label{tesla-landkarte}TESLA the region to the north--west of the DESY--Laboratory}
\end{figure}

Following up on this study a detailed plan for the TESLA site
north--west of the DESY--Laboratory has been worked out for this
Technical Design Report (see Fig.~\ref{tesla-landkarte}) and
preparatory planning work has been pursued in close co-operation with
authorities in the region. The tunnel of the linear accelerator starts
at the DESY site in a direction tangential to the straight section
West of HERA. The central research campus of the TESLA laboratory is
situated about 16\,km from the DESY site in a rural part of the North
German State (``Land'') of Schleswig-Holstein, and accommodates both
the collider detector hall for Particle Physics, and the FEL radiation
user facility.

In March 1998 the State Governments of Hamburg and Schleswig--Holstein
signed a treaty to jointly prepare all the necessary planning steps
and documents for the legal procedure (Planfeststellungsverfahren)
which can start immediately after the project authorisation. The
environmental aspects and the public participation are integral parts
of this procedure. The state treaty has been ratified by both state
parliaments.

\section{Radiation Safety}

The radiation safety requirements for the TESLA linear collider have been evaluated in detail. Different scenarios of beam losses were investigated. The radiation level has been minimised and is far below the natural dose.

The German regulations require that the maximum allowed personal dose due to direct radiation and radiation from radioactive release (activated air, water etc.) for the public does not exceed 1 mSv per year and for radiation from radioactive release alone to be less than 0.3 mSv per year. TESLA will stay below 1/10 of the above limits. This corresponds to a personal dose for the public of 0.1 mSv per year from direct radiation and radiation from radioactive release. In comparison, the natural doses in the northern part of Germany are about 1-2 mSv per year. 

Our studies of the radiological impact on the environment were evaluated by two independent German institutes. They confirmed the internal analysis and made some suggestions for improvement which were included in the technical design.

\section{General Safety Aspects} 

Safety of personnel and equipment in the tunnel has to be provided during construction, shutdown, maintenance, and operation of TESLA. The tunnel has segments between access shafts with a longest distance of 5\,km. The resulting escape and access times largely determine the organisational and technical means for rescue as well as for fire protection:
\begin{itemize} 
\item Access to the tunnel is restricted to instructed personnel.
\item Fire loads are minimised and kept in compartments or small hermetic containers acting as fire compartments.
\item Several levels of fire control and fire fighting are in place which react very early to keep fires well localised. The automated or remotely controlled levels foreseen include (1) component malfunction detection and early smoke detection, (2) cutting the electric power to components and (3) tunnel segments, (4) local fire control or suppression, and (5) remotely controlled fire fighting using the monorail trains. The final level includes human emergency intervention in the tunnel.
\end{itemize}
The experience gained with HERA in matter of safety is incorporated in the technical design. The technology available for tunnel safety is developing rapidly and TESLA will benefit from this development. Staff specially trained for rescue and for fire fighting in the tunnel as well as in the experimental halls will be made available.

\chapter{The Next Steps}

This Technical Design Report is the result of an intense collaboration
of many scientists from around the world. As shown by the scientific
and technical studies TESLA opens unique possibilities in particle
physics and for research with the X-ray FEL.

By building and operating the TESLA Test Facility and the associated
laser the TESLA collaboration has shown that the technology is at hand
to build a 500\GeV\  electron--positron collider with an integrated X-ray
laser. Results from an advanced R\&D on superconducting cavities
provide a credible path for an energy upgrade of the linear collider
towards a total energy of 800\GeV. The test facility also provides an
excellent basis for a reliable cost estimate, which has been worked
out in detail by  many experts.

The Deutsches Elektronen--Synchrotron DESY proposes to the international scientific community, to the German Federal Government and to the northern German state governments (``L\"ander'') to build TESLA in the vicinity of Hamburg.

In preparation of the German position concerning the approval of the
TESLA project the German Research Ministry has asked the German
Science Council (Wissenschaftsrat) to review TESLA together with other
large scale projects presently under consideration. The Science
Council is an independent body set up to advise the federal and state
governments of the Federal Republic of Germany on matters of higher
education and research policy. The planned large scale facilities
include besides TESLA the European Spallation Source and a Heavy Ion
Accelerator Facility.

In parallel to this evaluation a number of international review
processes are taking place on a European and world--wide scale,
addressing the long--term road map of particle physics, the scientific
potential of an electron--positron collider and of X-ray lasers, the
technologies for building linear 
colliders and XFELs, and models for building and operating large
international research facilities.

This in--depth scientific and technical report on TESLA will provide
the necessary input for these reviews.


\end{document}